\documentclass{cimento}

\usepackage{wrapfig}
\usepackage{graphicx}
\usepackage{comment}
\usepackage{relsize}

\usepackage{amssymb}
\usepackage{amsmath}

\newcommand\brabar{\raisebox{-4.0pt}{\scalebox{.15}{
\textbf{{\Big{(}}}}}\raisebox{-4.0pt}{{\_}}\raisebox{-4.0pt}{\scalebox{.15}{\textbf{\,{\Big{)}}
}}}}

\newcommand{\degree}{\ensuremath{{{}^\circ}}}

\newcommand{\mev}{\ensuremath{{\mathrm{MeV}}}}
\newcommand{\gev}{\ensuremath{{\mathrm{GeV}}}}

\newcommand{\second}{\ensuremath{{\mathrm{s}}}}
\newcommand{\gram}{\ensuremath{{\mathrm{g}}}}
\newcommand{\cm}{\ensuremath{{\mathrm{cm}}}}
\newcommand{\meter}{\ensuremath{{\mathrm{m}}}}
\newcommand{\meterPerSecond}{\ensuremath{{\mathrm{[m/s]}}}}
\newcommand{\kelvin}{\ensuremath{{\mathrm{K}}}}

\newcommand{\proton}{\ensuremath{{\mathrm{p}}}}
\newcommand{\neutron}{\ensuremath{{\mathrm{n}}}}

\newcommand{\eminus}{\ensuremath{{\mathrm{e}^{-}}}}
\newcommand{\electron}{\ensuremath{{\mathrm{e}}}}

\newcommand{\piminus}{\ensuremath{{\pi^{-}}}}

\newcommand{\barbracketnu}{\ensuremath{{\mathrm{\overset{\brabar}{\nu}}}}}

\renewcommand{\[}{\begin{equation}}
\renewcommand{\]}{\end{equation}}

\newenvironment{myitemize}
  { \begin{itemize}
    \addtolength{\itemsep}{-0.5\baselineskip}
    \addtolength{\baselineskip}{0\baselineskip} }
  { \end{itemize} }

  { \begin{enumerate}
    \addtolength{\itemsep}{-0.5\baselineskip}
    \addtolength{\baselineskip}{0\baselineskip} }
  { \end{enumerate} }

\shorttitle{Methods and problems in neutrino observatories}
\title{{\small  \textnormal{Proceedings ISAPP School ÒNeutrino Physics and Astrophysics,Ó 26 July - 5 August 2011,\\ Villa Monastero, Varenna, Lake Como, Italy}}
\\ \\ \\ \\

\noindent Methods and problems in neutrino observatories
}
\author{M.~Ribordy}
\instlist{\inst{} High Energy Physics Laboratory, EPFL, 1007 Lausanne, Switzerland}

\begin{document}


\thispagestyle{empty}

\maketitle

~\\ \\ \\ \\ \\ ~

\begin{abstract}
Gigantic neutrino telescopes are primarily designed to search for very high energy neutrino radiation from the cosmos. Neutrinos travel unhindered over cosmological distances and therefore carry unique undistorted information about its production sites: the most powerful accelerators of hadrons in nature.

In these lectures, we present the relevant physics motivations and their specifics. We review methodological aspects of neutrino telescopes: the experimental technique, some of the faced problems and the capabilities in terms of discovery potential, effective area, isolation of a signal from atmospheric backgrounds, etc. Instruments and their operation in various media are described. 
We also mention the instrumental birth and provide an outlook of the detection technique toward very low and ultra-high energies.

\end{abstract}

\newpage

\tableofcontents

\section{Introduction}
Following the discovery of cosmic rays (CR) in 1912, the field of elementary particle physics was founded upon the identification of a number of related particles such as positrons, muons, pions and strange particles. One hundred years later, however, the origins of the CR themselves remain obscure. The prevalent bottom-up paradigm posits their origin within cosmic accelerators, e.g. pulsar wind nebulae, X-ray binaries, supernova remnants in our galaxy or active galactic nuclei and gamma ray bursts, via the so called "Fermi acceleration" mechanism or alternatively in large scale electric fields present in pulsar environments. The interaction of the accelerated particles with the surrounding radiation fields or dense gas clouds in the neighborhood produces a secondary neutrino flux. This leads to the experimental possibility of conducting astronomy through the means of high energy (HE) ($\gtrsim\mathcal{O}(\text{10 GeV})$) neutrino messengers, perhaps providing the key to unlocking the mysteries of the CR origin. 

The nascent field of neutrino astronomy, at somewhat lower energies, has grown from the first successful observation of extra-terrestrial (ET) neutrinos from a supernova in 1987 and the sun, bearing witness to the death of a star and precipitating the monumental achievement of establishing a neutrino mass. 

The first neutrino telescopes in operation, NT200+ at Lake Baikal and AMANDA at the South Pole, attempted to open a new window on the Universe from the perspective of the HE neutrino, successfully detected HE neutrinos of atmospheric origin and have motivated the construction of larger detectors, the IceCube detector at the South Pole instrumenting a cubic kilometer of ice and Antares in the Mediterranean Sea, both in operation. 
As yet, however, no high energy extra-terrestrial signal has been detected, but phenomenological arguments suggest that signal detection should be possible with the completed IceCube array after a few years of data taking.

These detectors are not limited to the question of the CR origin and can be utilized to probe questions of fundamental particle physics, on the nature of the neutrino or on the validity of Nature's symmetries by means of the atmospheric HE neutrino beam, and cosmological scenarios. At the interface of these three pillars, fundamental particle physics, astrophysics and cosmology, HE neutrino telescopes may also be instrument in unveiling the nature of dark matter. 
This is discussed in Sections~2--6.

Section~7 introduces the detection technique and several methodological aspects: signatures, notions of effective areas and discovery potential, analytical approaches, event reconstruction and analysis systematics. 
Section~8 presents  historical elements in the experimental development of the field and its current status.

The choice of topics described in Sections~9--11 is intended to cover a broad spectrum of the methodology. Section~9 shows that high energy neutrino telescopes can be utilized as supernova detection instruments, i.e. they can probe a neutrino signal at energies many orders of magnitude below their original concept. A back of the envelope calculation on their potential is presented. Section~10 describes the generic procedure for point source  searches and introduces the important notions of upper limit, sensitivity and discovery flux. The results reported in recent years denote large methodological "breakthroughs" in the exploitation of  these instruments with a sophistication of the analysis methods. Section~11 concerns the ultra high energy detection methodology with the presentation of  the principle underlying alternative signatures.

In these lectures, we do not emphasize actual experimental results, which can be found in the mentioned literature.

\section{Cosmic Accelerators and neutrino astronomy}
The generic picture of the cosmic ray origin appears quite satisfactory up to $\approx 10^{20}$~eV to explain the CR energy spectrum and composition changes: CRs are accelerated in the vicinity of galactic and extra-galactic accelerators through efficient first order Fermi mechanism \cite{fermi-acc-pr}: particles, bouncing back and forth a strong shock propagating through interstellar plasma, gain energy  $\Delta E\propto E$ at each crossing of the wave-front with a small probability of escape of the region. This mechanism trivially predicts a power law spectral shape \[\mathrm{d}\Phi/\mathrm{d}E \sim E^{-\gamma}\] for the accelerated particles. Moreover, the spectral index takes a universal $\gamma=2$ for strong (supersonic) shocks with speed $\beta c$, due to the specific momentum imprint to the ISM from the shock crossing~\cite{longair,blandford-eichler}. Roughly, nuclei with atomic charge $Z$ will gain energy as long as they remain confined in the acceleration region of size $L$ and magnetic field $B$ ({\it i.e.} gyroradius smaller than $L$) and can therefore attain a maximum energy \[E_{\rm{max}}\sim Z\beta B L.\]

\begin{wrapfigure}{r}{6.5cm}
\centering\includegraphics*[scale=.28]{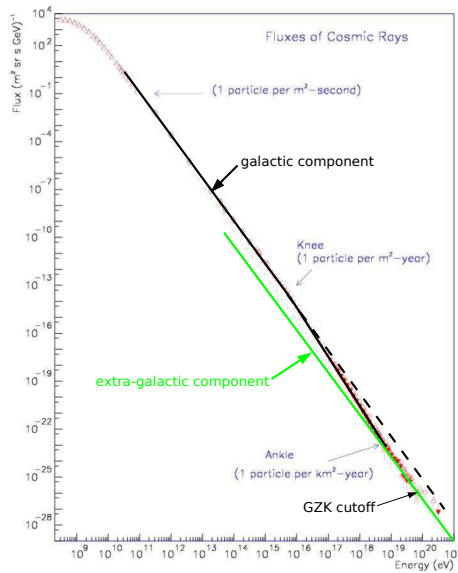}\\
\caption{The all-particle cosmic ray spectrum and its main features.}\label{fig:crspectrum}
\end{wrapfigure}

Several outstanding features characterize the CR spectrum illustrated Fig.~\ref{fig:crspectrum}: 
\begin{myitemize}
\item[-] A first break in the power law with a steepening of the CR spectrum occurs at a few PeV (knee). This break, which may sign the limitations of the confinement of the galaxy or galactic accelerators~\cite{cesarsky,Lagage:1983zz}, is accompanied by a gradual change of composition from light to heavy nuclei. 
\item[-] A second break occurs at a few EeV (ankle) and is understood as a cross-over of the fading galactic component and a harder extra-galactic component. This interpretation is further supported by a change of composition from heavy to light nuclei and by the fact that the galaxy cannot confine nuclei at these energies. 
CR's with energy exceeding that of the ankle are called ultra high energy (UHE) CR's.
\item[-] At ultimate energies, around $\approx 10^{20-21}$ eV per nucleon, protons interacting with the cosmic microwave background (CMB) undergo an energy damping due to photopion production over cosmologically short distances known as the GZK cutoff~\cite{gzk-cutoff}. See also Section~ref{sct:gzk} 
\end{myitemize}

Together with CR electrons, CR nuclei have provided valuable information in the study of prevailing acceleration processes and the morphology of the galaxy. However, the nature of cosmic accelerators is not revealed easily because CRs are poor astronomical messengers: deflected in magnetic fields, they do not  point back to their sources.

The occurrence of electronic acceleration (by the first order Fermi acceleration mechanism) up to about 100 TeV was demonstrated indirectly in supernova remnants (SNR) by means of conventional astronomy observing gamma radiation extending up to tens of TeV:
the accelerated electrons emit non thermal Bremsstrahlung radiation and synchrotron radiation, which in turn, together with other surrounding radiations become the targets for the inverse Compton process. The HE photons can then further produce electron pairs by interacting with the surrounding radiation fields. This dynamics produces complex spectral energy distributions (SED), typically with a double humped structure.

Following an old argument by Baade and Zwicky~\cite{baade}
the energy balance suggests that SNR's could account for the bulk of the galactic CR's (GCR) as well: given a CR density $\rho_{\rm{CR}}\approx1\,\text{eV/cm}^3$ and a CR galactic confinement time of about $\tau_{\rm{esc}}\approx10^7$ yr in a galactic volume $V_{\rm{gal}}\approx 10^{67}\,\text{cm}^3$, the CR luminosity is \[\mathcal{L}_{\rm{CR}}=\rho_{\rm{CR}}V_{\rm{gal}}/\tau_{\rm{esc}}\approx10^{41}\,\text{erg/s}.\]
With one supernova releasing $10^{51}$ erg every 30 yr ($10^9\,\text{s}$), $\mathcal{L}_{\rm{CR}}$ can be sustained provided a plausible conversion efficiency of about 10\% of the SNR energy into CR kinetic energy.

However, this postulate has  not been unambiguously established through multi-wave\-length (MWL) studies, because of the difficulty in disentangling the electronic component from the $\pi^0$-decay component signing hadronic acceleration (from processes $\mathrm{pp}, \mathrm{p}\gamma \rightarrow \pi \rightarrow \gamma, \nu$). The observation of HE neutrinos would provide an incontrovertible proof of hadronic acceleration.

\vskip3mm
Neutrinos as astronomical messengers present further advantages, as they may escape optically thick sources and provide an unrestricted access to the distant HE sky. Distant photons on the contrary interact with the extragalactic background light, producing electronic pairs and nearby blazar spectra are distorted if not absorbed above TeV energies\footnote{The situation is even more dramatic for PeV gammas interacting with the Cosmic Microwave Background (CMB):  if emitted in the galactic center, they would not reach us.}. However, neutrinos, with their small interaction cross sections are difficult to detect but searches for ET neutrinos are greatly alleviated by the combination of an increase of the cross section with growing energy and rather hard expected flux spectra $\mathrm{d}\Phi_\nu / \mathrm {d} E_\nu \propto E_\nu^{-2}$ as suggested by the first order Fermi acceleration mechanism. This is the reason for building large HE neutrino telescopes, in particular, IceCube~\cite{ic} and Antares~\cite{antares}, already mentioned, currently the most sensitive instruments in operation with complementary fields of views.
The discovery of a positive signal through the exploration of the neutrino sky would clearly represent a significant breakthrough and begin to resolve a hundred year old mystery. This is the key ambition of neutrino astronomy.

\vskip3mm
We note that conventional astronomy is continually revealing an ever richer sky at GeV and TeV energies, powered by astronomical objects. Among the favorite GCR accelerator candidates, besides Pulsar Wind Nebulae and Shell-like SNRs, which exhibit rather steady behavior,  we find other objects, notably associated to binary systems, which exhibit variable gamma-ray emission~\cite{lsi-magic, lsi-fermi, microq1,microq2,microq3,microq4,microq5}. 
Prime candidates to explain cosmic rays at the highest energies, beyond the ankle, are 
Active galactic nuclei (AGN), often displaying variability~\cite{Aharonian:2003hj} and  cataclysmic gamma ray burst (GRB) phenomena, which are quite common in the universe. The transient behaviors of some of these neutrino source candidates may facilitate the rejection of the atmospheric background and is therefore interesting from an experimental standpoint; this is particularly true for GRB's with analyses essentially background free because of the localized emission of a short burst of neutrinos, assuming the connection between  gamma-rays and neutrinos discussed below.

\subsection{UHE CR cosmic ray connection}
The energy density of the extra-galactic component extracted from the green line on Fig.~\ref{fig:crspectrum} is $\rho_{\rm{UHECR}}\approx 3\cdot10^{-19}\,\text{erg/cm}^3$. Therefore, the required power for a source population to generate $\rho_{\rm{UHECR}}$ over a Hubble time of $t_{\rm{H}}=10^{10}$ yr (per unit Mpc$^3$) is \[\mathcal{L}_{\rm{UHECR}}\approx 10^{45} \,\text{erg Mpc}^{-3}\,\text{yr}^{-1}.\]
From the perspective of astrophysics, the coincidence between this number and the 
electromagnetic output from AGN ($2\cdot10^{44}$ erg/s per AGN) and/or GRB ($2\cdot10^{52}$ erg per GRB) source populations, which fits the generic hypothesis of a transparent source, is the reason why these objects have come to be among the favored extra-galactic candidates for UHE hadronic acceleration.

In the transparent source, while accelerated protons remain trapped in the acceleration region and interact with the surrounding radiation field, the secondary neutrons and neutral decay products of secondary pions are no longer confined. 
This results in a similar energy injected into CR, $\gamma$ and $\nu$ escaping the source and related spectra:
the $\proton\gamma$ reaction proceeds through various modes, direct production, resonant (and multi-pions), \[\proton\gamma \rightarrow \pi^{+} \neutron,\, \pi^{0} \proton\]
and the neutrons, pions and muons further decay,
\begin{eqnarray}
&\neutron  &\rightarrow \piminus\proton\\
&\pi^{\pm} &\rightarrow \mu^{\pm} \barbracketnu_\mu  \rightarrow {\rm{e}}^\pm \nu_\mu \bar\nu_\mu \barbracketnu_{\rm{e}}\\
&\pi^0 &\rightarrow \gamma\gamma
\end{eqnarray}
Near the photo-pion production threshold, assumed to dominate, pions take away about 
$m_\pi/m_\proton$ energy fraction, while asymptotically pions take 50\%. Charged pions decay kinematics is distributing energy roughly a quarter to each final state particles. Eventually, the same amount of energy goes into gamma and neutrinos. For pp interaction, the picture brings to similar conclusion. Therefore, assuming a $E^{-2}$ neutrino spectrum,
($\int E_\nu \drm\Phi_\nu/\drm E_\nu=c\rho_{\text{\tiny UHECR}}/4\pi$), we have,
\[
 E^2_\nu \frac{\drm\Phi_\nu}{\drm E_\nu} \sim 5\cdot 10^{-8} \,\gev\, \cm^2\, \second^{-1}\, \text{sr}^{-1}.
\]
The diffuse neutrino flux from this generic transparent CR source is referred to as the Waxman-Bahcall (WB) upper bound \cite{wb}, once normalized on the UHECR component, and indicates the necessity to consider neutrino telescopes at the cubic kilometer scale (see Section~\ref{sct:meth}). 
This upper bound can be relaxed by assuming some optical opacity of the source, while maintaining consistency within experimental constraints (CR data and diffuse gamma-ray background). This is known as the Mannheim-Protheroe-Rachen (MPR) upper bound \cite{Mannheim:1998wp}.

We discuss now the blazar  as an example of hadronic acceleration to illustrate further the relation between gamma ray and neutrino emission.

\subsection{AGN blazar accelerator}

Blazars, an extreme manifestation of AGNs when the jets are pointing closely toward the observer are strong candidate sites for EG CR acceleration and for subsequent neutrino emission within hadronic models of activity given the compelling arguments for non thermal emission. Blazars come in two flavors, Flat Spectrum Radio Quasars (FSRQ) which shows broad emission line regions and low to high frequency BL Lac (LBL $\rightarrow$ HBL) characterized by the absence of broad emission lines and a smaller luminosities. A characteristic double humped spectrum is observed, which peaks in the infrared \& MeV bands for FSRQ and in the optical/UV - UV/X-ray \& GeV-TeV bands for LBL and HBL.
Many models of blazar activity predict the associated neutrino flux within reach of the current generation of instruments. They exhibit variability at all wavelengths and on various time scales (flares) which can be as short as one hour, constraining the size of the emission region to the submilli pc scale. A few blazars have shown remarkable and unusual behavior, emitting orphan flares, i.e. TeV outburst emissions with no counterpart at any other wavelengths \cite{orphan-obs}. This peculiar observation is hardly reconcilable with leptonic models of activity which predict a correlation between X-ray and TeV emission. Instead, it could sign the decay of mesons following p$\gamma$ interaction of accelerated hadrons, which then result in the emission of neutrinos~\cite{orphan} (the constraints on the correlated emissions are looser in hadronic models of activity). 

There are two broad categories of models of blazar activity, leptonic and hadronic and both can explain the double humped spectrum. A realistic model of $\gamma$-ray emission is likely to be the synthesis of both, if these objects are to be at the origin of a significant fraction of the extra-galactic CR flux.

Within the leptonic class of scenarios, protons are not accelerated at energies high enough for efficient p$\gamma$ interaction and the radiation is dominated by relativistic electrons. Synchrotron Self Compton (SSC) and Synchrotron External Compton (SEC) model classes are often considered in the literature~\cite{SSC}: external radiation, e.g. from the accretion disk, or synchrotron radiation within the jet of accelerated electrons constitute the target for the inverse Compton process in SEC resp. SSC models. 

Within the hadronic class of scenarios, when a significant fraction of the jet kinetic power is converted into the acceleration of relativistic protons, two main model classes are distinguished based on p$\gamma$ or pp interactions.

In the Synchrotron Proton Blazar model (SPB)~\cite{psb}, protons are co-accelerated with electrons but the energy density of electrons is small compared to the energy density carried by accelerated protons. However, the synchrotron emission by electrons accounts for a large fraction of the X-ray radiation.
The main TeV radiation process from proton synchrotron radiation is an essential ingredient in the reproduction of the double humped structure of the SED. Multiple processes initiate a HE $\gamma$ radiation, interaction of protons with the X-ray target, p$\gamma$ Bethe-Heitler pairs, p, $\pi$ and $\mu$ synchrotron losses and electrons from muon decay, inducing EM cascades, which redistribute the power to lower energies. Eventually, the emission region becomes optically thin and the radiation escapes. Contrary to the Proton Initiated Cascade model (PIC)~\cite{PIC},  which predicts large neutrino fluxes but is unable to reproduce the complex SED structure, very large magnetic fields are necessary in the SPB model and the resulting neutrino flux can be much lower. Experimentally, this model is interesting if periods of enhanced activity are considered (e.g. Mrk 501 in 1997) as the hardening of the spectra during these periods does not only translate into a larger neutrino flux but also into evolving spectral features.

Within the framework of pp scenarios~\cite{pp-blazar}, where accelerated protons interact with cold protons present in the jets, the observed double humped SED structure can be reproduced as well. In addition, given the low energy threshold of the reaction, this type of models requires only relatively low magnetic fields. The interaction rate can be high, resulting mainly in pions, which on decay produce neutrinos, the MeV--TeV radiation and further synchrotron emission from EM cascades may account for the observed X-ray radiation. The bolometric luminosity is related to the neutral pion component.

\begin{wrapfigure}{r}{6.5cm}
\centering\includegraphics*[scale=.3]{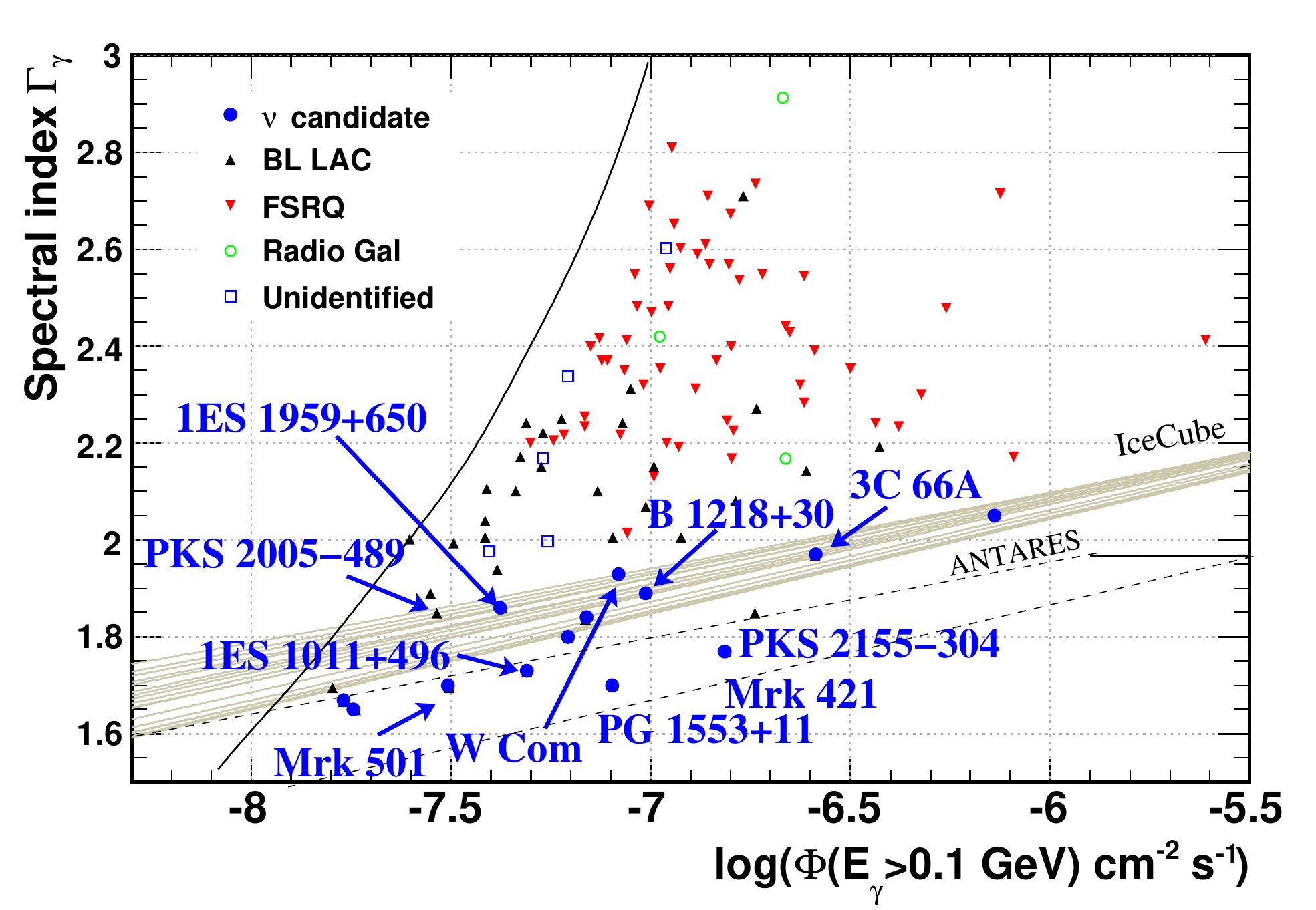}\\
\caption{From~\cite{ribordy-fermi}. Integrated flux $\Phi_\gamma$ w.r.t. the spectral index $\Gamma_\gamma$ distribution of {\it Fermi} AGN~\cite{abdo}. Superimposed, the gray curves represent the discovery threshold of IceCube in the pp$_{\rm{soft}}$ model (see text).}\label{fig:diagram}
\end{wrapfigure}

In the hadronic models of activity~\cite{ribordy-fermi}, the maximal flux of neutrinos allowed for pp and p$\gamma$ interactions was presented. In the case of an isotropic injection of soft protons (Fermi accelerated) and subsequent pp$_{\rm{soft}}$ interactions, the neutrino spectral shape follows closely the HE $\gamma$ spectral shape, as a result of source transparency to TeV radiation. Applied to the list of blazars characterized with {\it Fermi}~\cite{abdo}, the Fig.~\ref{fig:diagram} shows the discovery potential of IceCube, which was analytically calculated in~\cite{ribordy-fermi}, in a $\Phi_\gamma$ vs $\Gamma_\gamma$ diagram. Notice that the flux (integrated above 100 MeV) necessary for a detection drastically changes with the spectral index. Consequently, a few BL Lac would be within reach of IceCube but none of the bright FSRQs (due to softer emission spectra) often considered as favored neutrino source candidates in the literature.

Within this model, these results convey the importance of the necessity for a ``hybrid'' approach from the methodological standpoint in search for a neutrino emission, i.e. to consider the $\gamma$-ray integrated flux {\bf in combination} with the corresponding spectral index in order to optimally search for neutrinos. Moreover, assuming that an enhanced neutrino emission and a hardening of the neutrino spectrum may be expected from blazars in high states, this  prompts an analysis methodology, which relies on the time-dependence of the spectral shape and normalization of the neutrino emission, for optimizing the discovery potential.

In~\cite{ribordy-fermi}, p$\gamma$ and pp models with hard proton injection spectra were considered. In this case, the bolometric luminosity provides constraints on the upper neutrino flux level but the neutrino flux spectral shape, due to cascading processes of TeV photons, is not directly related to the GeV band spectrum. Therefore, in the eventuality of HE neutrino detection, the location of the detected sources in the diagram would strongly constrain the source acceleration mechanisms, opening the captivating prospects of delineating between hard (e.g. large scale electric fields at the base of the jet) and soft (first order Fermi acceleration) acceleration mechanisms.

\vskip3mm
We have presented above the basic motivations of neutrino astronomy and provided several hints calling for a generic point source search methodology. The reader will refer to the lectures by G.~Sigl, in this school, for a more detailed discussion on the phenomenology of high energy neutrino sources. We limit ourself to the methodological aspects in the following.

\section{Atmospheric neutrinos}
Along with  muons,  neutrinos are produced by the interaction of CRs with the atmosphere. 
Atmospheric neutrinos constitute an irreducible background in searches of ET neutrinos, contrary to atmospheric muons.
The atmospheric neutrino spectrum can be split into a steeply falling conventional component \cite{Volkova:1980sw,Volkova:1983yf,Honda:1995hz} and into a hard charmed or prompt component \cite{Lipari:1993hd,Costa:2001fb,Bugaev:1998bi,Gondolo:1995fq}. 

The conventional atmospheric neutrino component is relatively well known (uncertainties remain on the primary CR flux and the subsequent air shower modeling the kaon production) and results from the disintegration of pions and kaons, which progressively reinteract at growing energies. This "beam dump" effect  increments the spectral index by one unit ($\gamma=2.7 \rightarrow \alpha=3.7$)
\footnote{
In analogy to the spectral softening of electrons propagating in the galaxy,
the linear increase of the pion lifetime in the expanding shower  together with the linear increase of the energy loss per track length with energy results in an energy loss before decay in $E^2$, therefore the muon spectral index is $\alpha=\gamma+1$.}
This softening is essential in making the case of HE neutrino telescopes: eventually, there will be an energy above which a flux of ET neutrinos, assumed harder, will start to dominate. 
Kaon-induced neutrinos become increasingly dominant with energy and contribute already about 75\% of the neutrinos at TeV.

In contrast, the feeble prompt component remains hard up to PeV energies: the fast decay of very short lived charmed hadrons and mesons (produced as pairs D$\bar{\rm{D}}$, $\Lambda_c \text{D}$, etc.) prevents the "beam dump" to be effective up to energies well above PeV. Therefore, the prompt flux eventually surpasses the conventional flux above some critical energy, with a poorly estimated threshold around $0.1-1$ PeV due to the large uncertainties affecting its normalization from largely unknown cross sections ($\mathcal{O}(0.01-0.1\,\text{mb})$ in the range $0.1-1$ PeV).

The angular distribution of the conventional and the prompt components are qualitatively different:
below the critical energy, the absence of a beam dump for the short lived charmed mesons and hadrons results in an angular prompt neutrino distribution independent of $\cos\theta$, where $\theta$ is the zenith angle. For conventional mesons, the reinteraction rate is sensitive to the interplay between the decay length and interaction length, a function of $\cos\theta$: inclined showers  develop at first in comparatively low density atmosphere and therefore, the mesons  preferentially decay compared to vertical showers up to higher energies, therefore resulting in an increase of the HE neutrino flux. The flux suppression will depend linearly on $\cos\theta$, reflecting the slant depth behavior.

Therefore, an additional discrimination difficulty in searches for an isotropic diffuse flux of very high energy  (above PeV) neutrinos arises from the isotropic prompt neutrino flux component,  even if an ET neutrino flux is expected to be slightly harder.

\subsection{CR muon background}
The muon intensity decays exponentially with slant depth (the integrated column density along a path) penetration in the ground. At depths of about 14 kmwe, it turns into a constant: the muon flux is induced by neutrinos interacting in the neighborhood of the detector.
Neutrino telescopes,  projected or in operation, are located at depths between $1-5$ kmwe, i.e. at depths where the CR-induced muon flux is respectively $10^7 - 10^4$ times larger than the neutrino-induced muon flux. CR-induced muons constitute a reducible background to neutrino searches, in particular to atmospheric neutrino study, which depends on the  detector quality, i.e. its ability to reconstruct the incoming lepton direction. Thus, it is clear that burying the detector as deeply as possible helps to greatly alleviate this task. For instance, in order to isolate a sample of upward-moving atmospheric neutrinos with 90\% purity, the required discrimination power between downward muons and neutrinos is estimated at the $10^7$ level at a 2 kmwe depth.

\subsection{Strong ET neutrino discrimination}
HE  neutrinos events are conjointly produced  with muons in CR showers, opening the possibility of discrimination between downward-moving extra-terrestrial  and atmospheric neutrino events: the latter being accompanied by a muon (bundle), in particular a muon from the same parent meson. In~\cite{resconi-gaisser}, it was actually shown that
downward-moving atmospheric neutrino could be efficiently rejected in an IceCube-like detector with external VETO layers for zenith angles up to 60$^\circ$ and neutrinos with energy in excess of 10 TeV. 

\vskip3mm
The atmospheric neutrino beam is interesting on its own as a a signal for particle physics study (e.g. for charm production cross sections, neutrino oscillations, tests of the Lorentz violation, etc.). In principle it enables to picture the tomography of the earth~\cite{GonzalezGarcia:2007gg}.

\section{Neutrino oscillations}
Similarly to the quark sector, mass and weak eigenstates do not coincide in the neutrino sector. The unitary neutrino mass matrix $U_{\alpha i}$ relates mass to weak eigenstates according to $|\nu_\alpha\rangle =\sum_i U_{\alpha i} |\nu_i\rangle$. The evolution of a neutrino produced in a certain flavor state $\alpha$ is derived from the evolution of its mass eigenstates, i.e. governed by 
\[
|\nu_\alpha(x,t)\rangle = 
\sum_i {\rm{e}}^{-i ( E_i t - \bold{p}\cdot \bold{x})}\,U_{\alpha i} |\nu_i\rangle = 
\sum_{i,\beta} U_{\alpha i}\, {\rm{e}}^{-i(m_i^2/2E)t}\, U^T_{\beta i} |\nu_\beta\rangle 
\]
in the limit $\gamma\gg1$.
The propagated state differ in general from the initial state and the probability of appearance of a flavor $\beta$ at distance $L$ is
\[
p_{\nu_\alpha\rightarrow\nu_\beta}(L)=|\langle\nu_\beta|\nu_\alpha(L)\rangle|^2=\delta_{\alpha\beta}-4\sum_{j>i} \sin^2{(\delta m_{ij}^2 L/4E)}\, U_{\alpha i}\, U_{\beta i} \,U_{\alpha j} \,U_{\beta j}.\label{eq:posc}
\]
(considering real MNS matrix elements is correct for oscillations calculations in absence of CP violation).
Experimentally and in "good" approximation, $\{U_{ei}\}_i\approx\{c_{12},s_{12},0\}$, $\{U_{\mu i}\}_i\approx  \frac{1}{\sqrt{2}}\{-s_{12},c_{12},1\}$ and $\{U_{\tau i}\}_i\approx \frac{1}{\sqrt{2}}\{s_{12},-c_{12},1\}$, with $c_{12}=\cos\theta_{12}$ and $s_{12}=\sin\theta_{12}$, where $\theta_{12}\equiv\theta_{\rm{sol}}\approx 33^\circ$, assuming $\theta_{23}\equiv\theta_{\rm{atm}}\approx 45^\circ$ and $\theta_{13}\approx 0^\circ$; $\delta m^2_{21}\equiv\delta m^2_{\rm{sol}}\approx8\cdot10^{-5}\,\text{eV}^2$ and $|\delta m^2_{31}|\approx|\delta m^2_{32}|\equiv\delta m^2_{\rm{atm}}\approx2.4\cdot10^{-3}\,\text{eV}^2$ ($\delta m^2_{ij}=m_i^2-m_j^2$).

We leave it to the reader using eq.~\ref{eq:posc} to verify that the first dip in the neutrino oscillation probability $\nu_\mu\rightarrow\nu_\mu$ is at about 25 GeV for a neutrino crossing the earth through its center. Using the atmospheric neutrino beam, this region could be explored~\cite{Akhmedov:2008qt,Akhmedov:2006hb} provided a neutrino telescope with a sufficiently low energy threshold.

\subsection{Oscillations over astronomical baselines}
Distant astrophysical candidates of high energy neutrinos feature  $\delta m^2 L/E \gg 1$. This implies that the neutrino flux measured with a finite energy resolution $\Delta E/E$ is averaged over many periods of the oscillation probability argument $\sin^2{(\delta m^2 L/4E)}\rightarrow \int \drm E' \sin^2{(\delta m^2 L/4E')} g_\sigma(E'-E) = 1/2$, where $g_\sigma(E'-E)$ is a Gaussian distribution with width $\sigma=\Delta E/E$ for instance (moreover, the acceleration region is usually extended, of size $\Delta L$). One can therefore drop the dependences in $\delta m^2,\, E\,\text{and}\,L$ for the calculation of the neutrino flux at Earth,
\[
p_{\nu_\alpha\rightarrow\nu_\beta}^{\delta m^2 L/E\gg1}=\delta_{\alpha\beta}-2\sum_{j>i}  U_{\alpha i}\, U_{\beta i} \,U_{\alpha j} \,U_{\beta j}.
\]

Assuming the special $\theta_{23}$ and $\theta_{13}$ values and starting with a neutrino mixture at the production site \[\{\Phi^0_{\nu_\electron},\Phi^0_{\nu_\mu},\Phi^0_{\nu_\tau}\}=\{1,2,0\},\] which follows from both processes $\proton\gamma$ and pp (summing neutrinos and anti-neutrinos), it is straightforward to obtain the mixture propagated to earth  \[\{\Phi_{\nu_\electron},\Phi_{\nu_\mu},\Phi_{\nu_\tau}\}=\{1,1,1\}:\] all three flavors are equally populated at earth, opening the favorable experimental perspective of potential source detection with distinct signatures. Moreover, the presence of tau neutrinos potentially enables to look at up-going events in the very high energy regime, otherwise hindered.

If we would have kept the distinction between neutrinos and anti-neutrinos, we would have noticed that pp and p$\gamma$ leads to distinct mixtures. It is left as an exercise to conclude that the $\bar\nu_\electron$ flux is relatively lower in case of p$\gamma$, opening the experimental possibility of discriminating the dominant process at work in a cosmic accelerator via the Glashow resonance, provided it is detected at PeV energies (relative rates of track-like and cascade-like events are affected, see Section~\ref{sct:meth}).

\vskip3mm
The pattern of neutrino oscillation is affected in the presence of matter or assuming violation of the Lorentz invariance and briefly outline these topics below.

\subsection{Matter oscillations}
\begin{wrapfigure}{r}{6.5cm}
\centering
\includegraphics*[scale=.22]{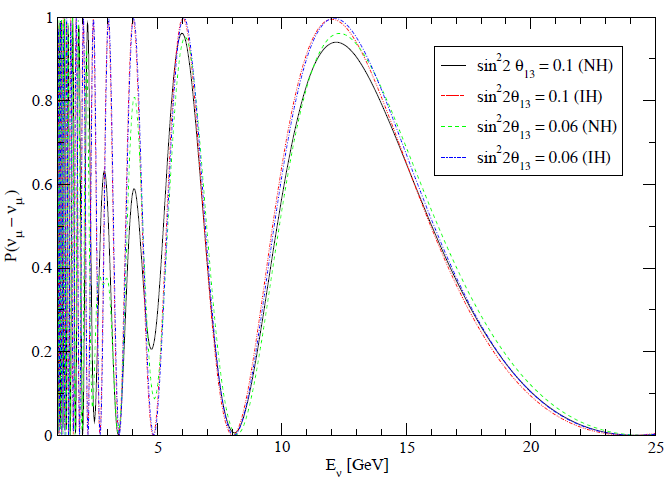}
\caption{$\nu_\mu\rightarrow \nu_\mu$ transition probability of upward-moving neutrinos for different $\theta_{13}$ values and hierarchy assumptions after earth crossing.}
\label{fig:hierarchy}
\end{wrapfigure}
In the presence of regular matter, the interaction of the electronic flavor differs from the other flavors. While in regular matter $\nu_\electron$ CC scattering on electrons is possible, this is forbidden for the other flavors and leads to the so-called Mikheyev - Smirnov - Wolfenstein (MSW) effect. In more formal terms, an additional potential appears in the Hamiltonian for the CC interaction and uniquely for the electronic flavor, so that mass eigenstates in vacuum are no longer mass eigenstates in matter. Working out the solution found in standard textbooks, we notice that the presence of matter eventually enables the disentanglement of the neutrino hierarchy for non vanishing $\sin^2\theta_{13}$ values (in vacuum we only have access to the absolute value of the difference of mass squared, not its sign). 

A neutrino telescope with a $\mathcal{O}(10\,\text{GeV})$ energy threshold may resolve the hierarchy, based on the statistical discrimination of $\nu$ from $\bar\nu$ (CC interaction cross sections differ by a factor two)\footnote{Neutrino telescopes cannot distinguish between the negatively and positively charged induced muons.}, provided rather large values of $\sin^2\theta_{13}\gtrsim 0.1$, as the oscillation pattern of neutrinos in the range between 10 and 15 GeV crossing the earth's core is most strongly affected by matter effects~\cite{Mena:2008rh}: matter effects enhance (or suppress) the oscillation probability of neutrinos or anti-neutrinos if the hierarchy is normal or resp. inverted (inverted resp. normal). 
Fig.~\ref{fig:hierarchy} illustrates the $\nu_\mu\rightarrow \nu_\mu$ transition probability of vertically up-going neutrinos for different $\sin^2\theta_{13}$ values and hierarchy assumptions after earth crossing (i.e. reaching the detector).

From an experimental point of view, resolving the neutrino hierarchy may turn out to be a difficult analysis dominated by systematic uncertainties:  good energy resolution and detection efficiency down to $E_\mu=5$ GeV are required (i.e. the ability of reconstructing tracks about 30~m long with about 10~m resolution), as well as a Monte Carlo providing relative flux systematic uncertainty  down to of a few percents w.r.t. the incoming neutrino direction on a moderately large angular scale (the kinematics enables only a neutrino incoming angle resolution of $\langle\Theta_{\nu\mu}\rangle\approx10\degree$ at these energies).

{\bf It is worth mentioning that,} in principle, the possibility to "distinguish" event by event between $\nu_\mu$ and $\bar\nu_\mu$ by assigning a tag probability exists. This, without measuring the  charge of the induced muon, but instead based on the precise topology of the contained event: the measurement of both, the energies of the induced muon (through its track length) and the energy of the induced shower at the interaction vertex, grants access not only to the incident neutrino energy but to the interaction inelasticity $y$ as well. The average value taken by $y$ differs for neutrinos and anti-neutrinos. Such an approach, in contrast to the statistical $\nu$ --  $\bar\nu$ discrimination based on cross section difference, will greatly help deciphering the neutrino hierarchy.

\subsection{Lorentz invariance violation}\label{sec:vli}
Many efforts to construct a quantum gravity~\cite{Gambini, Madore, kostelecky89} theory imply the deformation or the breaking of the Lorentz symmetry and predict non trivial modifications of space-time symmetries at the Planck scale, such as the existence of a new fundamental length scale $l_\mathrm{Pl}$. The Standard Model Extension framework~\cite{kostelecky98} provides an effective field-theoretical approach for the study of the violation of Lorentz invariance. Among the most promising experimental signatures~\cite{Mattingly,Stecker-vli,Coleman,hawking} within this phenomenological framework are neutrino oscillations as flavor changing signatures are amplified.

The modification of the dispersion relation, retaining the validity of the energy-momentum conservation law, is a simple kinematic framework to introduce a violation of the Lorentz invariance: \[E^2 = m_i^2 + (1+f_i)p^{2},\] where $i$ denotes the energy eigenstate and where the $\{f_i\}_i| f_i \ll 1$ depend on the species. 
At energies above 100 GeV (in order to  neglect mass-induced oscillation), the oscillation probability, left as an exercice, is \[p_{\nu_\mu \rightarrow \nu_\mu} = 1-\sin^2{2\theta_v}\sin^2{(\delta v EL/2)},\]
where $\theta_v$ is the maximum attainable velocities (MAV) mixing angle to the flavor eigenstates and $\delta v$ the fractional MAV difference. 
We note that (1) oscillations also occur if the neutrino mass vanishes and (2) the $EL$ dependence of the argument instead of the usual $L/E$ in standard oscillations. The exotic term competes with the classical term and may dominate at high energy,
leading to distortions of the angular and energy distributions relative to pure mass-induced oscillations. Distribution discrepancies beyond statistical and systematic errors would suggest a violation of the Lorentz invariance. Alternatively, this would help constrain the allowed parameter space ($\delta v$, $\theta_v$, etc.).
 IceCube in its baseline configuration will accumulate a unique sample of $\mathcal{O}(10^6)$ atmospheric neutrinos with energies above 50 GeV, representing an increase in sensitivity for testing a violation by more than an order of magnitude on existing constraints~\cite{Gonzalez-Garcia:2004wg,Battistoni:2005gy,amanda-vli}, in the particular case of coinciding mass and asymptotic velocity eigenstates~\cite{Gonzalez-Garcia:2005xw}.
A violation could be probed up to a fractional velocity difference approaching $\delta v=10^{-28}$, the limiting factor being the shrinking size of the neutrino sample above a growing energy threshold.

\section{GZK neutrinos}\label{sct:gzk}
UHE protons may interact with low energy photons (cosmic microwave or infrared background) $\proton\gamma \rightarrow \pi\neutron$, provided the CoM energy of the reaction satisfies $s\ge (m_\proton+m_\pi)^2$. This
corresponds to an energy threshold of $\approx 10^{20}$~eV for a frontal collision and a mean CMB photon energy of $\langle \epsilon \rangle = 6.4\cdot 10^{-4}$~eV and results in the degradation of the nucleon energy and the appearance of the so-called GZK cutoff~\cite{gzk-cutoff}.

Following the photo-pion production reaction, neutral and charged pions and neutrons are produced, which further decay into gamma rays and neutrinos.

The interaction length is approximately given by
\[\lambda_{\rm{GZK}}=(n_\gamma \sigma_{\proton\gamma})^{-1} \approx 3\,\rm{Mpc},\]
where $\sigma_{\proton\gamma}=0.25$~mb and $n_\gamma=411$~cm$^{-3}$.

Accounting more precisely for the exact shape of the photo-pion production cross section $\sigma_{\proton\gamma}(E_\proton, \epsilon_\gamma, \cos{\theta})$ with direct, multi-pion, diffractive and resonance contributions, where $\cos{\theta}$ is the angle between the proton and photon in the lab frame, and for the CMB spectrum $f(\epsilon_\gamma)$, the effective threshold is about $E_\proton\approx 10^{19.6}$~eV.
The interaction length is given by
\[\lambda^{-1}_{\rm{int}}(E_\proton) =  \frac{1}{2} \int \drm\cos{\theta}  \int_0^{\infty} \drm\epsilon_\gamma \sigma_{\proton\gamma}(E_\proton, \epsilon_\gamma, \cos{\theta})
f(\epsilon_\gamma) \]
and the attenuation length approximately (restricted to two-body final state) by including an inelasticity factor $K_p(E_\proton, \epsilon_\gamma, \cos{\theta})\equiv 1-E'_\proton/E_\proton$ in the integrand, which is about $1/8$ at the reaction threshold and $1/2$ asymptotically.

The attenuation length sharply decreases with increasing energies up to about $10^{20}$~eV due to the rapid increase of the inelasticity factor and cross section. Follows a slower decrease up to about $10^{20.5}$~eV due to the inelasticity increase, partly compensated by a decline of the cross section. It increases again  above $10^{21}$~eV because of the decrease of the cross section ($K_\proton$ is saturated).

The relatively short interaction length ensures that GZK neutrinos are astronomical messengers keeping track of the original CR direction. Should there exist a few UHE cosmic accelerators located close-by ($\lesssim$Gpc), GZK neutrino detection would allow the possibility of pinpointing the location and determining the nature of the most powerful cosmic ray accelerators in the universe.

\begin{wrapfigure}{r}{5.5cm}
\centering
\includegraphics*[scale=.4]{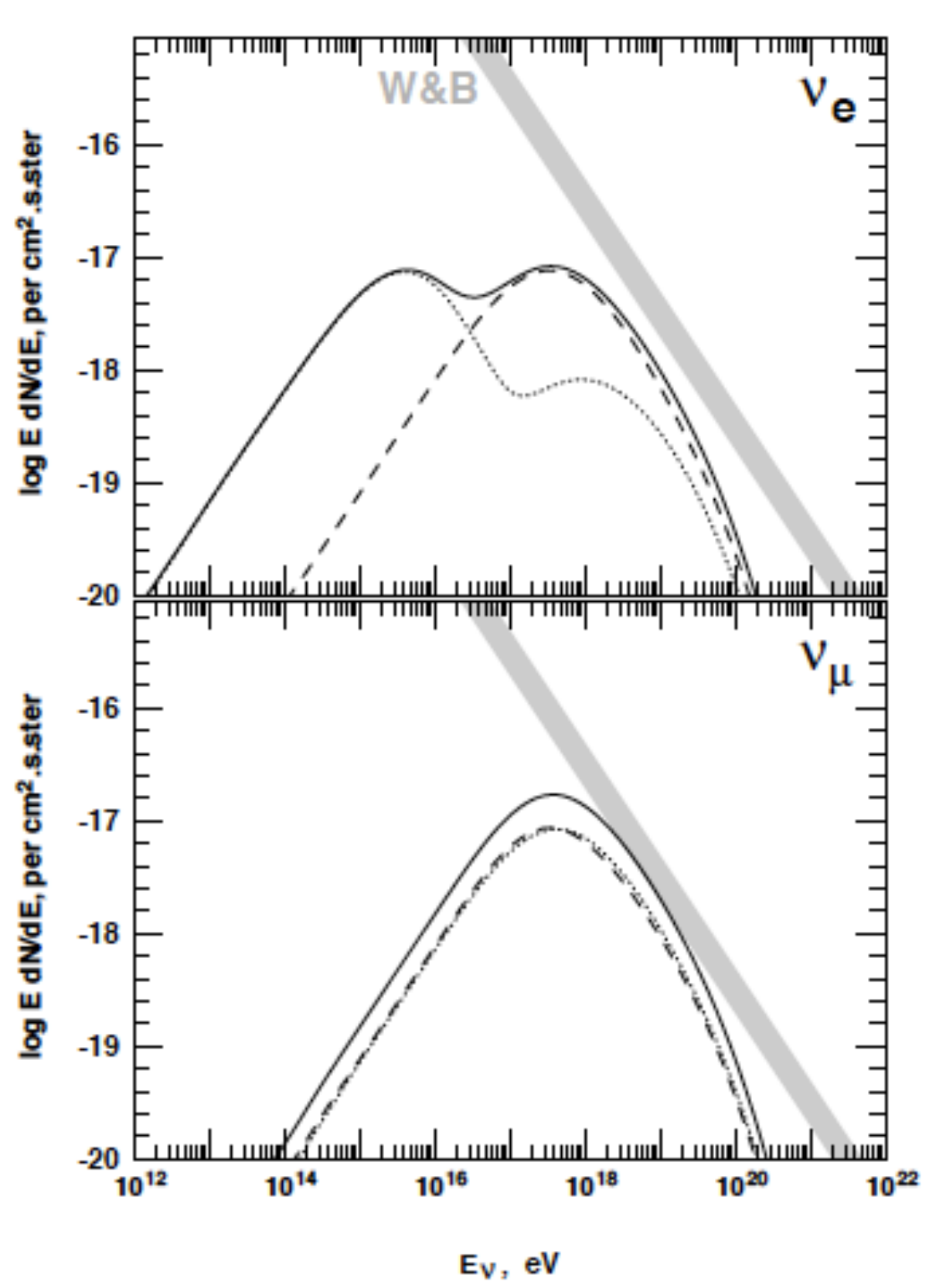}
\caption{GZK neutrino flux. Fig. from~\cite{Engel:2001hd}.}
\label{fig:gzkflux}
\end{wrapfigure}
The GZK neutrino flux has two distinct components, reflected by a double hump structure of the expected GZK neutrino flux, illustrated Fig.~\ref{fig:gzkflux}, which arises from the kinematic consideration of meson decay at the higher energies around $10^{18}$~eV (contribution to $\nu_{\rm{e}}, \nu_\mu, \bar\nu_\mu$ and a bit of $\bar\nu_{\rm{e}}$) and from the neutron decay at the lower energies around $10^{16}$~eV (contribution to $\bar\nu_{\rm{e}}$):
while neutrinos from meson decays carries about 1/30 of the proton energy (at reaction threshold), neutrinos from neutron decay carries about 100 times less energy in average than from meson decay.

The exact shape of the GZK neutrino flux depends on  CR composition~\cite{Anchordoqui:2007fi}, cosmological UHE CR source evolution and the injection spectrum of these CR, which is often characterized by a power law and the injection cutoff energy.

These parameters, essential to precisely predict the GZK neutrino flux, can be partly constrained from the CR spectral features~\cite{Seckel:2001mv}. However, the energy of CRs is degraded upon propagation and a fraction of the CR information is lost and could be partly recovered with a  characterization of the neutrino flux~\cite{seckel-stanev}.
We understand the necessity of a multi-messenger approach, combining neutrinos and CR information.

Heavier nuclei are photo-dissociated along their path~\cite{Hooper:2006tn}. Currently, the situation is uncertain: the existence of nuclei with $E\gtrsim10^{20}$ eV is questioned, there may be rising evidences for a gradually heavier composition above $\approx 10^{19}$ eV~\cite{AUGER,Kampert:2009zz}, compatible with an intrinsic limitation of the highest injection energy and/or the GZK mechanism on the one hand. On the other hand, a possible correlation of UHE CR sources with the AGN distribution by AUGER hints at a light composition~\cite{Abraham:2007si}.

Detailed theoretical aspects of this topic are discussed in the lectures by G. Sigl, in these lecture proceedings.

\section{Dark matter}
Under certain assumptions, the Minimal Supersymmetric Standard Model  framework provides a stable weakly interacting dark matter (DM) candidate: the neutralino, a self-annihilating thermal relic of the early universe. Its mass, bounded from below by accelerator constraints and above by theory, lies between 46 GeV~\cite{abda} up to a few TeV~\cite{Griest:2000kj}. Secondary particles, including $\nu$ (directly or indirectly), are emitted at a higher rate from regions of greater DM density, where gravitationally trapped neutralinos annihilate pairwise~\cite{carlos}
(neutralinos in the galaxy do pairwise annihilate with a rate which is proportional the square of the density). 
The galactic halo or compact objects such as the sun seem to be promising regions of such enhanced DM densities and thus for conducting dedicated analyses for these signatures, which all share a common approach searching for an excess from the directions of these enhanced DM density regions.
In the absence of any excess over the known atmospheric 
neutrino background, upper limits on the neutrino-induced muon flux from DM annihilations are obtained.
The deeper connection to the physics arises with the conversion of the neutrino-induced muon flux limits into cross section upper limits~\cite{Wikstrom:2009kw}: 
self-annihilation cross sections, velocity averaged (in halo analyses), $\langle\sigma_A v\rangle$, and spin-dependent scattering cross sections $\sigma_{\mathrm{SD}} $ in search for signatures from the self-annihilating solar neutralinos (assuming equilibrium between capture and annihilation rate in the Sun).
Assuming neutralinos constitute a sizable fraction of the galactic DM density, these analyses have a significant potential to exclude regions from the MSSM parameter space which would otherwise remain unconstrained by direct search experiments~\cite{Halzen:2009vu} and by $\gamma$-ray and CR experiments performing indirect searches similar to IceCube~\cite{Abbasi:2009uz}.

\subsection{Solar WIMPs}
We repeat the old but enlightening derivation of an estimate the number of events in a neutrino telescope following~\cite{halzen-stanev-gaisser}: a WIMP halo density $\rho=0.4\,\gev \,c^{-2}\,\cm^{-3}$ in corotation with the galaxy $v_\chi=300\,\text{km/s}$ is necessary to explain observed galactic rotation curves, leading to corresponding number density and flux
\begin{eqnarray}
n_\chi &=& 8\cdot10^{-4}\left(\frac{500\,\gev/c^2}{m_\chi} \right) \,\cm^{-3}\\
\Phi_\chi &=& n_\chi v_\chi = 2\cdot10^{4}\left(\frac{500\,\gev/c^2}{m_\chi} \right) \,\cm^{-2}\,\text{s}^{-1}
\end{eqnarray}

Assuming $\sigma_{\chi \rm{N}} = (G_F m_{\rm{N}}^2)^2/m_{\rm{Z}}^2$, we can then derive the solar WIMP capture cross section and corresponding rate,
\begin{eqnarray}
\sigma_{\rm{sun}} &=& \frac{m_{\rm{sun}}}{m_{\rm{N}}} = 1.2\cdot10^{57} \times 0.5\cdot10^{-41}\,\cm^2\\
\Gamma_{\rm{cap}} &=& \Phi_\chi \sigma_{\rm{sun}} = 1.2\cdot10^{20}\,\text{s}^{-1}
\end{eqnarray}
for $m_\chi=500\,\gev /c^2$.

To finally calculate the rate of solar neutrino of dark matter origin annihilating in the center of the sun, we assume a steady state with capture and annihilation rates of WIMPS are in equilibrium (the sun has traveled the galaxy multiple times), i.e.
$\Gamma_{\rm{ann}} =\Gamma_{\rm{cap}} /2$. The dominant annihilation channel is into weak bosons, each producing muon neutrinos with a branching ratio around 10\%, $\chi\bar\chi\rightarrow\text{WW}\rightarrow\mu\nu_\mu$, each carrying about half of the neutralino energy. The neutrino generation rate is therefore related to the capture rate $\Gamma_\nu=\Gamma_{\rm{cap}}/10=1.2\cdot10^{19}\,\text{s}^{-1}$ and the flux at earth is given by
\[
\Phi_\nu=\frac{\Gamma_\nu}{4\pi d^2} = 0.5\cdot10^{-8}\,\cm^{-2}\,\text{s}^{-1},
\]
where $d=1$ a.u.

Considering roughly half of the neutrino energy is transferred to the muon (i.e. $E_\mu=100\,\gev$, interaction inelasticity is approximatively 50\%), the neutrino-induced muon event rate in neutrino telescope with a cross section $A=1$~km$^2$ is therefore given by
\[
N_{\rm{event}}=A \, \Phi_\nu \rho_{\rm{ice}} \sigma_{\nu\rightarrow\mu}R_\mu \approx 20\,\text{yr}^{-1},
\]
where $\sigma_{\nu\rightarrow\mu}=0.7\cdot10^{-36}\,\cm^2$ and muon range $R_\mu=300$~m.

This is a very small number considering the large atmospheric neutrino background at these energies, about 2--3 per year and per square degree and the rather large kinematic angle between the neutrino and the induced muon, of order 5 degree. Such a signal could eventually be isolated as an excess over background provided a good detector efficiency after  several years. However, for larger WIMP masses, the potential will slowly degrade due to the decreasing statistics partly compensated by the fading atmospheric neutrino background. For smaller WIMP masses, it will degrade rapidly due to a number of unfavorable factors: increasing atmospheric background, decreasing detection efficiency (muon range decrease) and interaction cross section.

\subsection{Neutralino annihilation in the galaxy}
The flux of secondary neutrinos, which travel toward the detector from a given direction can be calculated by integrating the WIMP annihilation rate along the line of sight \[J(\Psi)\propto \int\drm l \rho^2(r(l,\Psi)).\] This approach is model-dependent because the dark matter density profile is not known \cite{DMprofile1,DMprofile2,DMprofile3}. It is believed to be cuspy at the center and rather flat far from it (where it is less model-dependent). Obviously therefore, the high rate would be from the direction of the galactic center. Unfortunately, this is a region of high background for the largest operating neutrino telescope IceCube located at the South Pole. However the potential is not lost, the large background rate being partly balanced by the high expected rate~\cite{Abbasi:2011eq}.

The neutrino flux is~\cite{Yuksel:2007ac}
\[\frac{\drm\Phi_\nu}{\drm E}(\Psi)\propto\frac{\langle\sigma_A v \rangle}{2} \, J(\Psi) \, \frac{\drm N_\nu}{\drm E}.\]
where $\langle\sigma_A v \rangle$ is the thermally averaged annihilation cross section, which is the averaging of cross-section times velocity, on the random velocity distribution, the factor 2 because the neutralinos self-annihilate and $\drm N_\nu/\drm E$  is the neutrino multiplicity on self-annihilation of a neutralino pair. The WIMP miracle comes from the fact that the annihilation cross-section is of the order of the weak cross-section if we require them to explain the dark matter (from freeze out arguments in big bang cosmology).

\section{Neutrino telescope detection methodology}\label{sct:meth}
Neutrino telescopes are instruments  optimized for the detection of upward-going neutrino above 10 -- 100 GeV, using the earth to filter other particle species (muons), which originate in CR showers.
The astronomical practicability relies on the quickly fading irreducible atmospheric neutrino background with energy compared to expected harder ET neutrino fluxes.
Current instruments can observe various signatures with limited flavor discrimination power and  consist of photomultipliers installed in a  medium (water or ice) transparent to the Cherenkov light emitted by relativistic particles induced by a neutrino interaction. The contamination by downward-going atmospheric muons is reduced by placing the detector deeper under ice/water and instrumenting it more densely. Typically, the distance between the PMT's will be chosen according to the low energy threshold one wants to achieve and the medium optical characteristics.

\subsection{Neutrino interaction cross sections}
The neutrino-induced muon interaction cross section $\sigma_{\barbracketnu{\rm{N}}}\equiv\sigma(\barbracketnu N\rightarrow \mu^\pm X)$, where $N=\frac{1}{2}(n+p)$ is an isoscalar nucleon, above a few GeV is~\cite{gandhi1,gandhi2,Strumia:2006db}
\[
\frac{\drm^2 \sigma(\barbracketnu N\rightarrow \mu^\pm X)}{\drm x \drm y} = \frac{2 G_F^2 m_{\rm{N}} E_\nu}{\pi} \,\left( \frac{m_{\rm{W}}^2}{Q^2+m_{\rm{W}}^2} \right)^2\, (x q(x,Q^2)+x \bar q (x,Q^2) (1-y^2))
\]
written in terms of the  Bjorken scaling variables: $x=Q^2/ 2 m_{\rm{N}} y E_\nu$ is the fraction of momentum carried by the struck quark ($-Q^2$ is the 4-momentum transfer between initial neutrino and outgoing lepton) and $y=1-E_\mu/E_\nu$ is the fraction of neutrino energy transferred to the hadronic system ($y$ is called inelasticity factor). $G_F$ is the Fermi constant and $q(x,Q^2)$ are the quark distribution functions, which have been measured: $\barbracketnu_\mu$ interaction proceeds with valence quark ${\rm{d}}_v$ (${\rm{u}}_v$) and sea quark $\bar{\rm{u}}_s$ ($\bar{\rm{d}}_s$) at leading order. 

For illustration, in the range of energy between GeV up to a few TeV (i.e as long as $E_\nu\ll m_{\rm{W}}^2/2m_{\rm{N}}$) can be simply expressed as~\cite{stanev-gaisser}
\[\drm\sigma_{\nu{\rm{N}}}/\drm E_\mu (E_\nu,E_\mu) \approx (0.72+0.06(E_\mu/E_\nu)^2) 10^{-38}\,\cm^2 \,\gev^{-1},\] and with slightly different factor for $\bar\nu_\mu$.
Therefore (left as a trivial exercice), the total cross section 
\[\sigma_{\nu{\rm{N}}}(E_\nu)= \int_0^{E_\nu} \frac{\drm\sigma_{\nu{\rm{N}}}}{\drm E_\mu} (E_\nu,E_\mu)\drm E_\mu\] 
behaves linearly with energy and the fraction of the energy transferred in average to the lepton is 
\[\frac{\langle E_\mu\rangle}{E_\nu} \equiv 1- \langle y\rangle = \frac{1}{E_\nu}\frac{1}{\sigma_{\nu{\rm{N}}}} \int_0^{E_\nu} E_\mu \frac{\drm\sigma_{\nu{\rm{N}}}}{\drm E_\mu} (E_\nu,E_\mu)  \drm E_\mu\] is about 50\%.
At higher energies, we should account for deviations from the W boson propagator. Above 0.1 PeV, the cross section is increasing logarithmically, $\sigma_{\nu\proton}(E_\nu)\propto E_\nu^{0.4}$,   and the inelasticity is decreasing (larger energy fraction transferred to the lepton) and reach about $y=0.2$ above PeV.

The extrapolation of neutrino cross sections becomes more undependable in ultra high energy interactions, which emphasize small $x\lesssim10^{-4}$ unprobed regions, together with dominant theoretical uncertainty arising from $x$ extrapolation to small values.
Within the standard model, the current knowledge of the quark structure functions ensures their accurate computations up to about 100 PeV~\cite{arXiv:1102.0691,subir1,subir2}.

Neutrino interaction cross sections with nucleons dominate over the ones with electrons, due to $m_\proton\gg m_\electron$, except in the region around $E_\nu=m_{\rm{W}}^2/2m_\electron \approx 6.3$~PeV, corresponding to the resonant W boson production \[\bar\nu_\electron\eminus\rightarrow {\rm{W^{-}}} \rightarrow \bar\nu_l l^{-},\, \text{hadrons}\] (the Glashow resonance): the W boson decay into hadrons (68\%) and equally in each lepton (total 32\%) and the interaction cross section in the resonance region is about \[\sigma(\bar\nu_\electron\eminus\rightarrow W)\approx 0.5\cdot10^{-30}\,\cm^2.\]

Neutrinos traveling through the earth begins to be absorbed above TeV, due to the rise of the cross sections. Modeling the earth density according to the Preliminary Reference Earth Model (PREM)~\cite{prem}, the integrated column density $X(\theta)$ along a chord w.r.t. the incidence zenith angle $\theta$ can be calculated. The probability of transmission of neutrinos is given by
\[
p_{\rm{tr}}(E_\nu,\theta) = {\rm{e}}^{-\sigma_{\nu{\rm{N}}}/m_{\rm{p}}\int \drm l\, \rho(l)} =  {\rm{e}}^{-X(\theta)/\Lambda_{\rm{int}}(E_\nu)}
\]
where the interaction length (in g/cm$^2$) is $\Lambda_{\rm{int}} = 
m_\proton/\sigma_{\nu{\rm{N}}}$.

The average transmission probability $\langle p_{\rm{tr}}(E_\nu) \rangle_\theta$ of an isotropic upward neutrino flux (referred to as the shadowing factor $S(E_\nu)$) is
\[
S(E_\nu)\equiv\langle p_{\rm{tr}} (E_\nu)\rangle_\theta =  \int_{-1}^{0}{\mathrm{d}}\cos{\theta}\, {\rm{e}}^{-X(\theta)/\Lambda_{\rm{int}}(E_\nu)}
\]
The case for tau neutrinos is slightly different as they may regenerate when crossing the earth~\cite{tau-regen}, see Section~\ref{sct:tau}.

\begin{figure}
\includegraphics[width=0.48\linewidth]{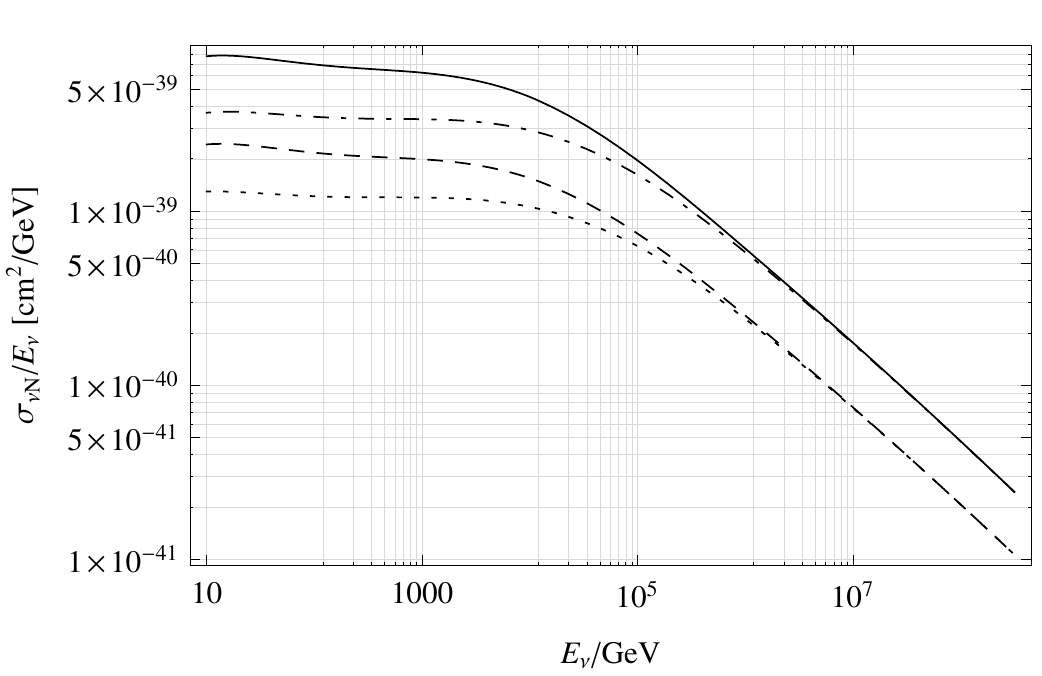}~\hfill
\includegraphics[width=0.48\linewidth]{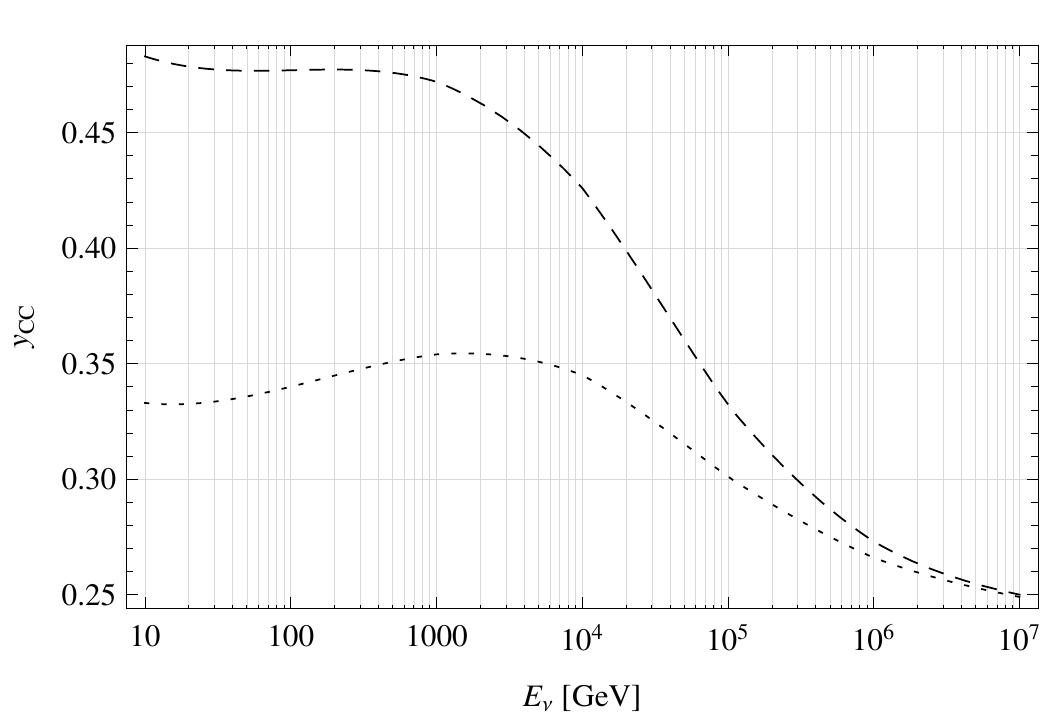}
\includegraphics[width=0.48\linewidth]{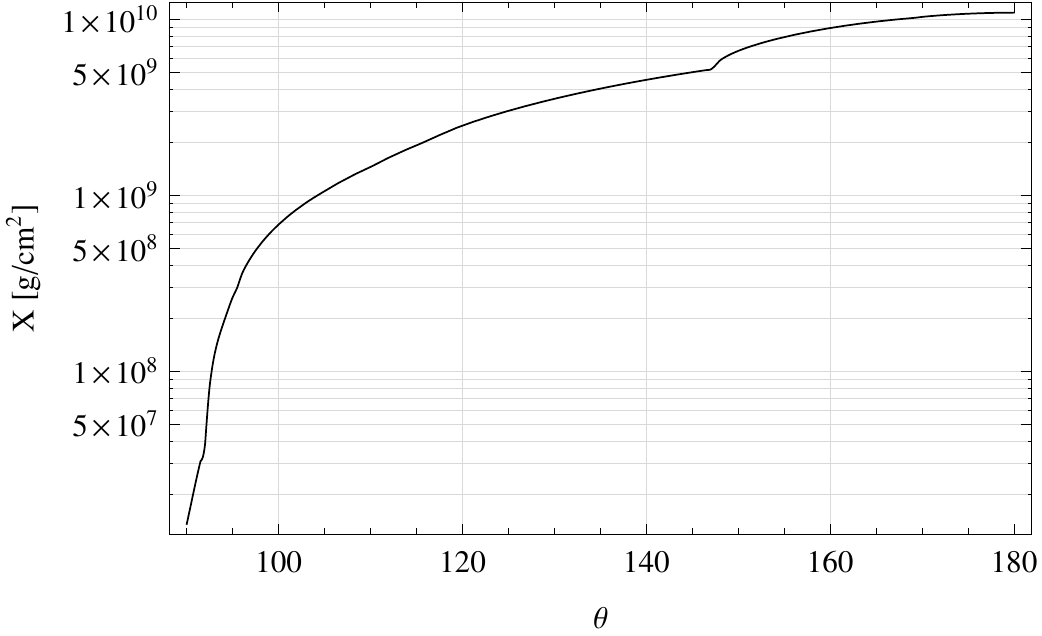}~\hfill
\includegraphics[width=0.48\linewidth]{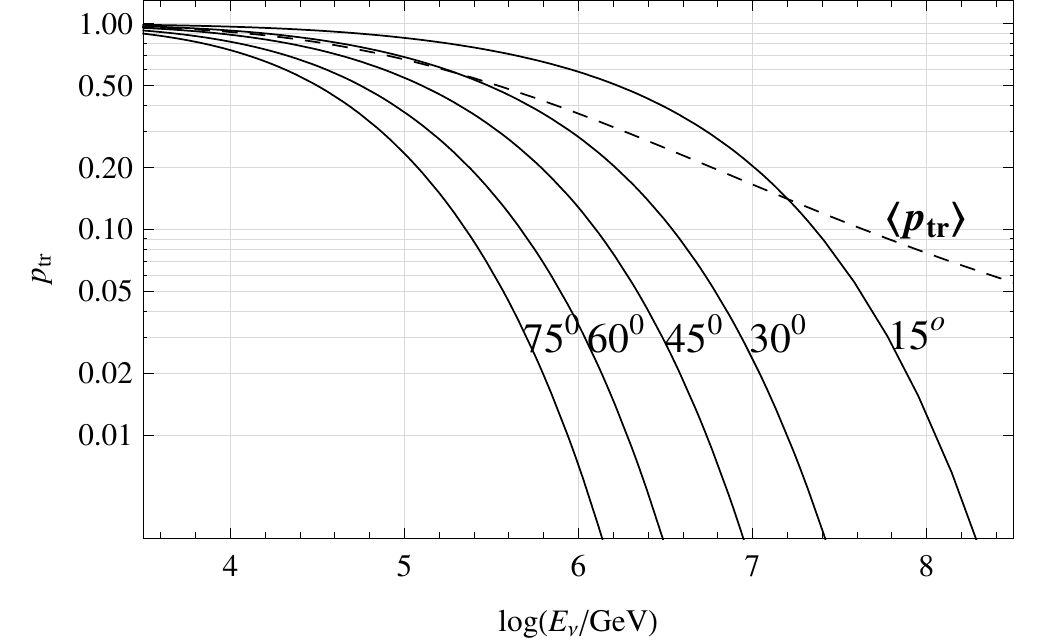}
\caption{Top left: $\nu_\mu$ cross sections for CC $\nu_\mu$ (solid), NC $\nu_\mu$ (dashed), CC $\bar\nu_\mu$ (dashed-dotted) and NC $\bar\nu_\mu$ (dotted). Top right: inelasticity parameter $y_{\rm{CC}}$ for muon neutrinos (dashed) and anti-neutrinos (dotted). Bottom left: Earth integrated column density $X$ w.r.t. the zenith penetration angle. Bottom right: transmission probability of $\nu_\mu$ for various zenith angle and averaged for an isotropic diffuse flux.}
\label{fig:numuXsec} 
\end{figure}

\subsection{Event topologies}
All flavors interact with nucleons through neutral (NC) and charged current (CC) interactions. Charged current interactions with electrons are forbidden for $\bar\nu_{\mu,\tau}$. NC interactions and $\nu_\electron$ CC interaction end up producing shower-like signatures referred to as cascades. Muon neutrino CC interactions initiate a shower and a muon, resulting into a cascade and long track-like signature. The short lifetime of the charged tau lepton and its decay channels will usually only produce a single resolvable cascade; only with energies $\gtrsim$PeV, the track-like signature may be distinguished ($\tau$ track length of about 50 m/PeV), leaving distinct signatures such as double bang (both the initial $\nu_\tau$ interaction shower and the final shower initiated by the $\tau$ decay are visible) and (inverted) lollipop (only one of the shower is occurring in the detector).

\begin{figure}[ht]
\centering
\includegraphics*[scale=.3]{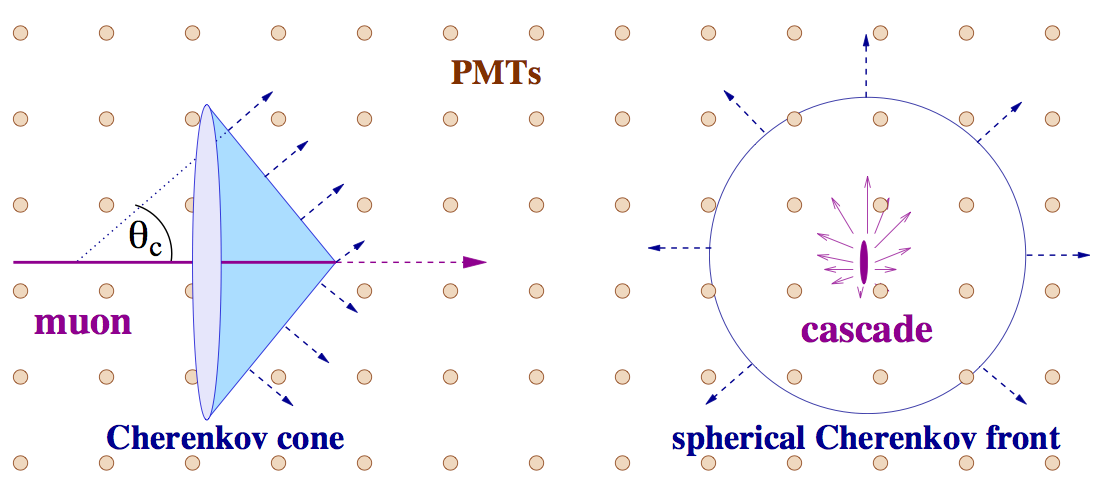}
\caption{Shower and track detection modes. Fig. from~\cite{Ahrens:2003fg}}
\label{fig:detectionModes}
\end{figure}

\paragraph{Muon neutrino}
\begin{wrapfigure}{r}{6.5cm}
\centering
\includegraphics*[scale=.45]{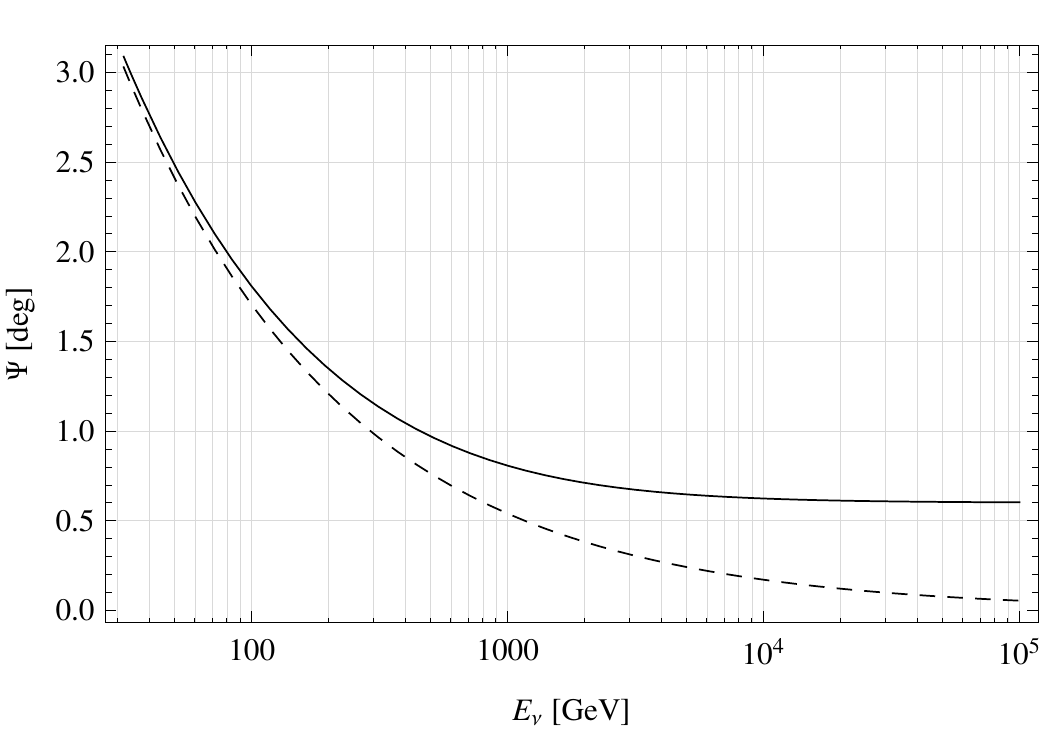}
\caption{Estimated IceCube $\nu_\mu$ angular resolution. Dashed: kinematic component.}
\label{fig:angRes}
\end{wrapfigure}
At energies above 10--100 GeV, neutrino-induced muons is the golden channel to search for point sources of ET neutrinos in neutrino telescopes: 

(1) CC interactions result in long muon tracks crossing the detector, collinear with the parent neutrino: the average angle between $\nu_\mu$ and $\mu$ from the kinematics is $\Theta_{\nu\mu}(E_\nu) \approx 0.6^\circ / \sqrt{E_\nu/{\rm TeV}}$ and the long leverage enables to achieve resolutions better than a degree as shown Fig.~\ref{fig:angRes} and therefore provide a mean for astronomy. In contrast, the angular resolution of cascade-like events is less favorable.

(2) Muon neutrinos can be detected whenever the Cherenkov light emitted along the induced muon track is recorded by the sensors. Combined with the fact that a muon track length surpasses the extension of current detectors above about 300 GeV, the interaction does not have to necessarily happen within the instrumented volume contrary to cascade-like event. This feature significantly boost the detection potential of $\nu_\mu$.


For these reasons, we have and will continue to focus mainly on this channel:
these features combined with the increasing cross section with greater energies maintain some constant "detectability" for a power law neutrino spectrum $\drm\Phi/\drm E_\nu\propto E^{-\gamma}$, with spectral index $\gamma=2$, up to about 10 TeV. Above the "detectability" begins slowly decreasing because of significant earth absorption and slower increases (instead of linear) of the muon range and cross section.

The muon energy loss can be parametrized 
\begin{equation}
\frac{\drm E_\mu}{\drm X}=-(a+bE_\mu),
\label{eq:mueloss}
\end{equation}
where $X$ is the  amount of matter traversed by the muon on its way to the detector and  
$a\approx 2.23\,\mev/(\gram/\cm^2),\,b\approx4.63\cdot10^{-6}/(\gram/\cm^2)$ are the nearly energy independent standard rock coefficients~\cite{mmc} (of course these coefficients also change with the medium, which eventually turns into water or ice). 
The critical energy $\epsilon=a/b\approx 500\,\gev$ marks the transition between continuous  and stochastic energy losses. The latter are due to interactions of the energetic muon with the (Coulombian field of the) nucleus, pair producing pairs or disrupting it and bremsstrahlung. Overall, pair production and bremsstrahlung processes dominate the muon nuclear interaction. These are of extreme catastrophic nature and typically, a 100 TeV muon will loose a significant fraction of its energy within only a few stochastic processes. 

Integrating expression (\ref{eq:mueloss}), we obtain the amount of matter, which is crossed  in average by  a muon of initial energy $E_{\mu_0}$ and final energy $E_\mu$,
\begin{equation}
X(E_\mu,E_{\mu_0})=\frac{1}{b}\ln{\frac{E_{\mu_0}+\epsilon}{E_{\mu}+\epsilon}}.
\end{equation}

$R_\mu(E_{\mu_0}) \equiv X(0,E_{\mu_0})$ is called the muon range. While at low energy the range increases linearly, the range only increases logarithmically when stochastic losses dominate. However, the range can become very large (the range of a 1 PeV muon in ice is $R_\mu\approx20$~km), much larger than the extension of a detector.

Showers produced along the muon track through the stochastic processes are producing additional light, which makes the precise reconstruction of the direction of the muon more difficult and the estimation of the crossing muon energy more precise (continuous emission is not providing any hint to the energy of the muon).

\vskip3mm
At energies well above PeV, the sky is opening to downward-going muons, due to a low atmospheric background, an interesting possibility given earth absorption largely prevents the detection of upward-going neutrinos.

\paragraph{Tau neutrino}\label{sct:tau}
Upon $\nu_\tau$ CC interaction, a large fraction of its energy is transferred to a $\tau$ lepton with a very short lifetime. The $\tau$ will decay back before loosing a significant amount of energy most of the time into a $\nu_\tau$ neutrino (either directly or through hadronic decays), which keeps in average 20\% of the initial energy.
Upon NC interaction,  $E_{\nu_\tau}$ is reduced by a factor two in average.
Therefore, a $\nu_\tau$ flux is not absorbed so dramatically as the other flavors when penetrating earth at VHE~\cite{tau-regen}. 
This feature, called tau neutrino regeneration, ensures the experimental possibility of  the observation of neutrino point sources at VHE, given the equal mixture of all neutrino flavors from cosmic accelerators after propagation toward earth (however, the original energy has been degraded and this effect only marginally improves the detection potential of VHE neutrinos).

Also, the tau decay branching ratio of into a muon is $\approx 18$\%. Again, given an equal  mixture of all neutrino flavors from cosmic accelerators after propagation over long baseline distances, the $\nu_\mu$ point source search potential at moderate energies, up to 10--100 TeV, is slightly enhanced (we are actually only recovering a fraction of the muon neutrinos which have been oscillated into $\nu_\tau$ during propagation).

\subsection{Effective area, event rate and detection potential}
Given a detector with a section $A$ and length $L$, assuming $R_\mu\gg L$, a muon propagating normal to $A$ toward the detector will be detected if the neutrino interaction has occurred within a distance $<R_\mu$ from the instrumented volume. The effective volume can therefore be estimated $V_{\rm{eff}}=AR_\mu$ (for cascade-like events, in contrast, $V_{\rm{eff}}=AL$ translating the fact that these events are detected if they occur within the instrumented volume).

The probability of interaction in the effective volume is approximately given by $p_{\nu\rightarrow\mu}(E_\nu)=R_\mu(\langle E_{\mu_0}\rangle)/\lambda_\nu(E_\nu)$, where $\lambda_\nu(E_\nu)$ is the interaction length of a neutrino with energy $E_\nu$ ($\Lambda_{\rm{int}}(E_\nu)\equiv\rho\lambda_\nu(E_\nu)$) and 
$\langle E_{\mu_0}\rangle$  is obtained from $E_\nu$ by means of the interaction inelasticity discussed above.

The number of detected events in a time $T$ from a monochromatic neutrino flux $\Phi_\nu(E_\nu)$ can now be roughly estimated
\[N=TAp_{\nu\rightarrow\mu}(E_\nu)\Phi_\nu(E_\nu).\]
The quantity $Ap_{\nu\rightarrow\mu}(E_\nu, \langle E_\mu\rangle)$ is called the neutrino effective area $A^\nu_{\rm{eff}}$ (the neutrino flux damping term $p_{\rm{tr}}$ is neglected in this illustration). 

\vskip3mm
A less crude approximation of $p_{\nu\rightarrow\mu}$ can be made integrating over the differential cross section and setting a muon energy detection threshold $E_\mu$,
\[p_{\nu\rightarrow\mu}(E_\nu, E_\mu)=\frac{1}{m_\proton}\int_{E_\mu}^{E_\nu}\drm E'_\mu
\frac{\drm\sigma_\nu}{\drm E'_\mu}(E_\nu,E'_\mu)X(E_\mu,E'_\mu).
\]
We obtain for an arbitrary neutrino flux at earth,
\[N=TA\int_{E_\mu}^{E_\nu}\drm E_\nu  p_{\nu\rightarrow\mu}(E_\nu,E_\mu) \frac{\drm\Phi_\nu}{\drm E_\nu}(E_\nu) {\rm{e}}^{-X(\theta)/\Lambda_{\rm{int}}(E_\nu)}.\]

Together with the angular resolution function, the effective area is an essential notion to describe the potential of a detector. Its knowledge enables to immediately calculate the expected number of events from a  source, point-like or diffuse.
We generically have
\[
N=T\sum_{\alpha=\electron,\mu,\tau} \int_{E_{\nu_\alpha}} \drm E_{\nu_\alpha} A^{\nu_\alpha}_{\rm{eff}}(E_\nu,E_\mu,\theta) \frac{\drm\Phi_{\nu_\alpha}}{\drm E_{\nu_\alpha}}(E_\nu)
\]
for a point source with flux $\drm\Phi_{\nu_\alpha}/\drm E_{\nu_\alpha}$ (for a diffuse source, an integration over the solid angle additionally appear).
Keep in mind that in order to keep the notation light, we have dropped the time dependence of the zenith angle $\theta$ and time integration: in general, neutrinos from a point source traveling to a detector will penetrate earth with $\theta(t)$, but for a detector located at a pole.

\vskip3mm
At this point, however, considering the notion of neutrino effective area only, we are missing an essential point (we concentrate on muons) because we can only count neutrinos from a given model. But  how do we define the search area $\Omega$ to compute the background in order to optimize the sensitivity when searching for ET neutrinos from a neutrino point source and how do we assess whether two models can be distinguished ? A neutrino telescope measures the muon energy / incoming direction and not the neutrino energy / incoming direction after all. For this,

\vskip1mm
(1) The knowledge of the muon effective area $A_{\rm{eff}}^\mu(E_\mu,\theta)$, a geometrically ambiguous and ill-defined notion, which can be extracted from 
\begin{equation}
\label{eq:effareas}
{\int {\rm{d}}E_\nu A_{\rm{eff}}^\nu(E_\nu,\theta) \frac{{\rm{d}}\Phi_\nu(E_\nu)}{{\rm{d}}E_\nu}
= \int {\rm{d}}E_\mu A_{\rm{eff}}^\mu(E_\mu,\theta) \frac{{\rm{d}}\Phi_\mu(E_\mu,\theta)}{{\rm{d}}E_\mu}.}
\end{equation}
where the differential muon flux is calculated via the propagation of the muon flux from the interaction vertex as
\[\label{Fmu}
\frac{{\rm{d}}\Phi_\mu(E_\mu,\theta)}{{\rm{d}}E_\mu} = \int {\rm{d}}E_\nu p_{\rm{det}}(E_\mu,E_\nu)  p_{\rm{tr}}(E_\nu,\theta)p_{\rm{int}}(E_\nu) \frac{{\rm{d}}\Phi_\nu(E_\nu)}{{\rm{d}}E_\nu}.
\]
$p_{\rm{det}}(E_\mu,E_\nu)$ is the probability density for a muon produced with energy $E_{\mu_0}(E_\nu)$ to reach the detector with energy $E_\mu$,
\[p_{\rm{det}}(E_\mu,E_\nu) = \frac{-{\rm{d}}X(E_\mu,E_{\mu_0}(E_\nu))/{\rm{d}}E_\mu}{R_\mu(E_{\mu_0}(E_\nu))}
=  \frac{1}{\ln{(1+E_{\mu_0}(E_\nu)/\epsilon)}}\frac{1}{(E_\mu+\epsilon)}\] and set to zero outside of the interval $E_\mu\in[0,E_{\mu_0}(E_\nu)]$.

\begin{figure} 
\centering
\includegraphics[width=0.45\linewidth]{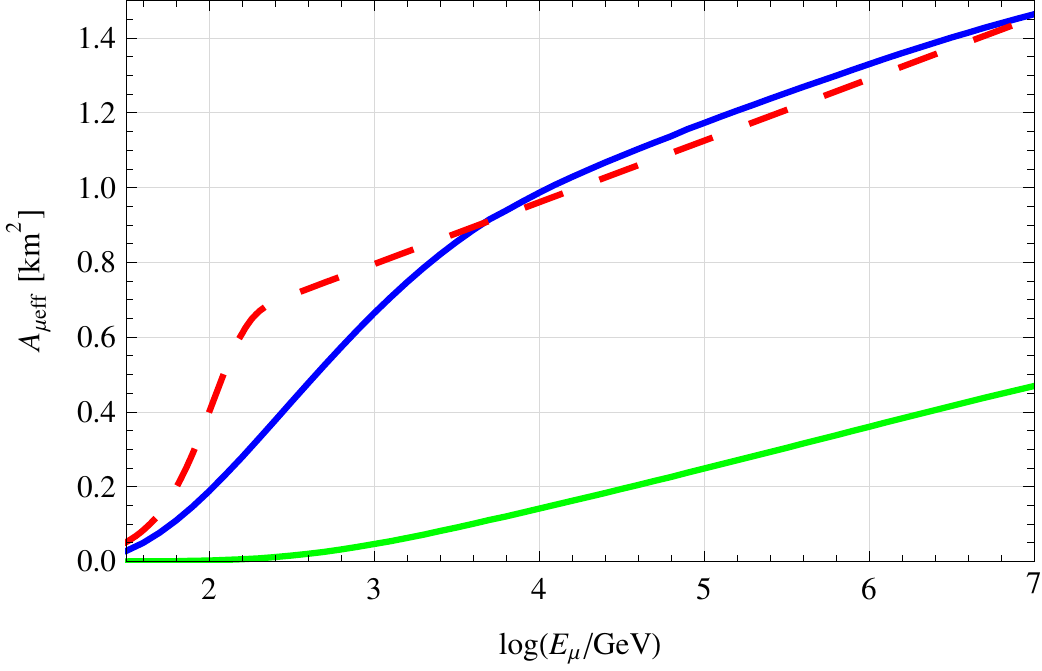}
\includegraphics[width=0.45\linewidth]{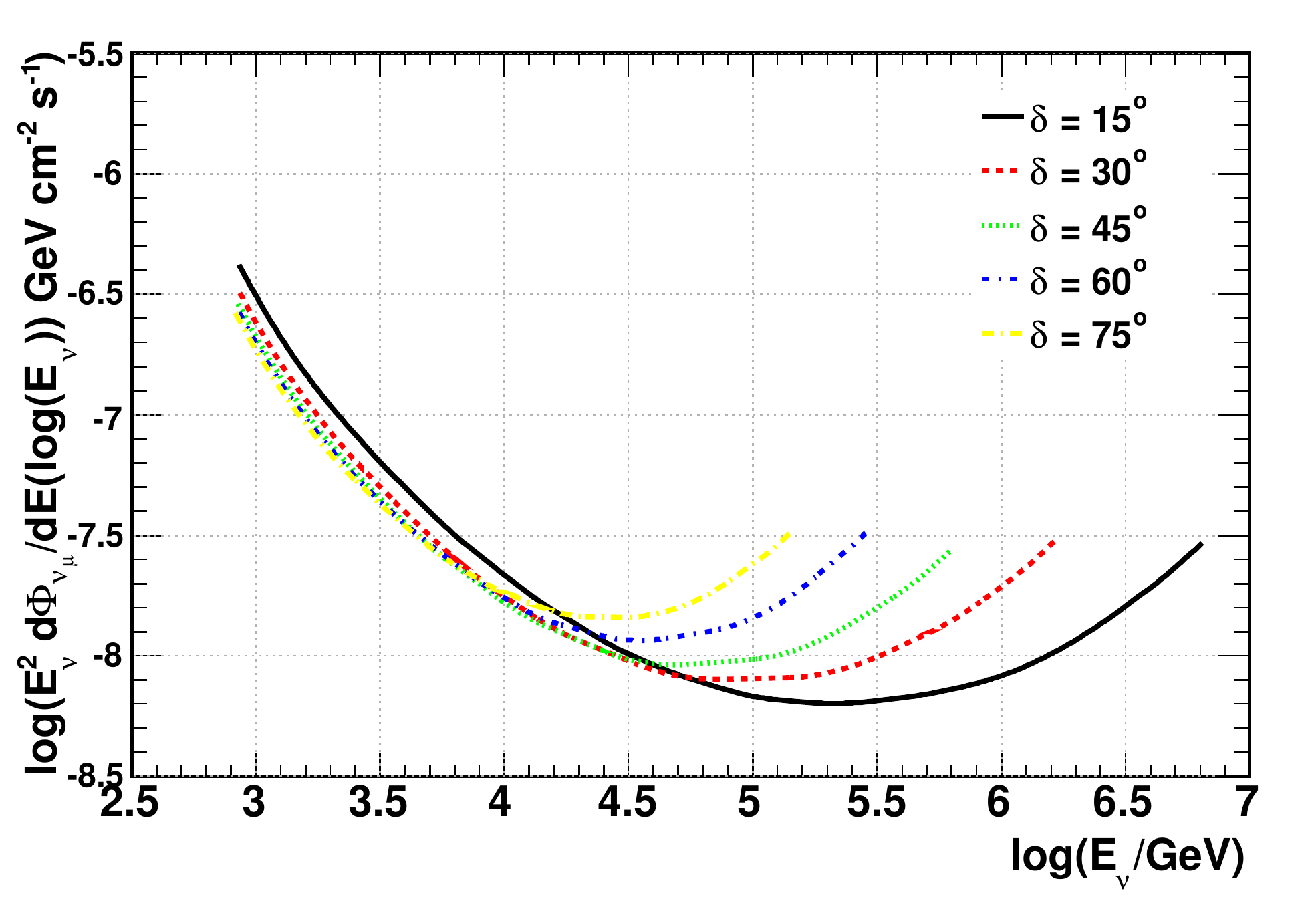}
\caption{Left: Full lines are the extracted muon effective areas for IceCube, when in construction (22 strings) and completed (80 strings). The dashed curve is the muon effective area derived in~\cite{Gonzalez-Garcia:2005xw}. Fig. from~\cite{microq5}. Right: IceCube point source reach after three years of exposure. The tangent to one of these curves intersecting the y-axis at 1 GeV defines the muon neutrino flux normalization for the corresponding declination. Fig. from~\cite{ribordy-fermi}.}
\label{fig:nu_spectranu_aeff} 
\end{figure}

\noindent $p_{\rm int}(E_\nu)=N_A \sigma_{\rm{CC}}(E_\nu) R_\mu(E_{\mu_0})$ is the neutrino interaction probability in the vicinity of the detector (potentially producing a muon within the reach of the detector) and $\sigma_{\rm CC}$ and $\sigma$ are respectively the charged current and the total muon neutrino cross sections.
The muon effective area, extracted according to the recipee found in~\cite{microq5} from averaged neutrino effective areas found in~\cite{nuEffMontaruli} is shown on the left of Fig.~\ref{fig:nu_spectranu_aeff}.

\vskip1mm
(2) The knowledge of the detector energy and angular resolution functions is necessary in order to optimize the search area for signal to noise maximization (best detection potential) and to calculate the {\it reconstructed} differential muon spectrum.

In~\cite{ribordy-fermi}, we have shown how to extend the analytical treatment above and thus extract as well discovery curves, which have been found {\it a posteriori} to be in remarkable agreement with discovery curves following a full analysis of the IceCube data.
They are shown on the right of Fig.~\ref{fig:nu_spectranu_aeff} for various source declinations and bear the following meaning: 
a neutrino flux with spectral index $\Gamma_\nu$ (corresponding to a straight line in the $\log(E)$ vs. $E^2 \log(\mbox{ Flux})$ representation) tangent to one of the spectral discovery curve is at the limit of discovery.
This representation of the IceCube discovery limit(s) is useful, because it enables
\begin{enumerate} 
\item The estimate of the sensitivity for arbitrary assumptions about the slope of the neutrino spectrum (i.e. the normalization of the minimal detectable spectrum at a given reference energy)
\item The estimate of the neutrino energy range contributing most significantly to the source signal (the energy at which the minimal detectable source spectrum grazes the discovery curve).
\end{enumerate}
The dependence of  the discovery curves on the source location in the sky is reduced to the source declination  $\delta$, at a polar location.
We notice a weak dependence of the discovery limit on the source declination in the case of soft neutrino sources ($\Gamma_\nu \gg 2$): Most events contributing to the source signal have relatively low energies, at which earth is transparent for neutrinos. 
For sources with hard spectra, $\Gamma_\nu \ll 2$, the  discovery potential is strongly affected by the increase of the source declination. The rise of the earth absorption probability with energy obscures the source signal, contributed mostly by high energy neutrinos.
The effect is most dramatic for the hardest neutrino spectra, $\Gamma_\nu\simeq 1$. In this case, the minimal detectable flux for the sources at the declination $\delta=75^\circ$ is about 1.5 orders of magnitude higher than from sources at $\delta=15^\circ$.


Experimentally, all of the complications above vanish and the potential of the detector and its power to reject models or delineate concurrent model comes straightforward from the analysis, relying partly on the detector Monte Carlo (mainly for the simulation of the signal).

\subsection{Event reconstruction}
Tracks or showers are reconstructed using a probability density distribution (PDF), which depends on measured time and location of the detected Cherenkov photons (hits). 

Consider a Cherenkov light source (e.g. a track or a shower), which can be parametrized by giving a vertex $\vec{q}$,  incidence angles $\theta$, $\phi$ and its energy $E$, that is $n_\mathrm{d.o.f}=6$. For a source with a different topology, e.g. an extended source, the parameters should be defined accordingly and $n_\mathrm{d.o.f}$ acquires a corresponding value.

The reconstruction of the parameters consists in maximizing the log likelihood expression for an arbitrary hit time series for each receiver~\cite{Ahrens:2003fg},
\begin{equation}
\label{llhE}
\ln{\cal L} = \frac{1}{n_\mathrm{hit}-n_\mathrm{d.o.f}}  \Bigg(\sum_{j=1}^M \Big(
\ln{p^{j}_{N_j}} 
+ \ln{f^{j}_{N_j}} \Big)
+ \sum_{\{j|N_j=0\}} \ln{f^{j}_0}\Bigg)
\end{equation}
where $j \in \{1,\,..,\,M\}$ is the hit receiver index, $n_\mathrm{hit}=\sum_j N_j$ is total number of hits in the event,
 $p^{j}_{N_j}\equiv p^{j}(t_1,\, ...\, t_{N_j})\propto \prod_{i=1}^{N_j}p_i^j$ is the PDF of detecting $N_j$ photons at receiver $j$ and at times $\{ t_1,\, ...\, t_{N_j} \}$ ($p_i^j$ is the PDF of detecting a photon at receiver $j$ and at time $t_i$) and $f^{j}_n$ is the probability to detect $n$ photons at receiver $j$, assumed to obey Poisson law $f^{j}_n = e^{-\mu}\mu^n/n!$, where $\mu \equiv\mu^{j}$ is the mean number of detected photons.
$p^{j}_i$'s and $\mu$'s depend implicitly on the relative orientation and distance between the source and the receiver, the optical properties of the medium, the receiver efficiency, etc.
The $\mu$'s additionally depend on the source intensity (related to the source energy).
Some detail on the specific expressions of these PDF's as used in IceCube can be found in~\cite{conv-ribordy}.
The last term represents the information from receivers which have not detected any photons. 

We notice that eq.~(\ref{llhE}) can be split in two parts depending separately on energy $E$ and directional $\{\vec{q},\,\theta,\,\phi\}$ parameters. This leads to introduce the reduced log likelihood formulas:
\begin{equation}
\label{llh}
\ln{\cal L}_\mathrm{dir} = \frac{1}{n_\mathrm{hit}-n_\mathrm{d.o.f}} \sum_{j=1}^M 
\ln{p^{j}_{N_j}} 
\end{equation}
and
\begin{equation}
\label{llh_onlyE}
\ln{\cal L}_E = \frac{1}{n_\mathrm{hit}-n_\mathrm{d.o.f}}  \Bigg(\sum_{j=1}^M
\ln{f^{j}_{N_j}} + \sum_{\{j|N_j=0\}} \ln{f^{j}_0}\Bigg)
\end{equation}
where $n_\mathrm{d.o.f}$ take the corresponding values (respectively 5 and 1). 

Expressions~(\ref{llh}) and~(\ref{llh_onlyE}) can be used separately to reconstruct the corresponding subset of parameters in a first approximation, when calculation speed or convergence is a concern. This will consequently be accompanied by a degradation in the precision of the reconstructed parameters and  best results are obtained by using expression~(\ref{llhE}), which more fully exploits the available information.

A more tractable, hybrid, so called MPE (multi photo-electrons), reconstruction, which leads to very good reconstruction results is based  on $\ln{\cal L}_\mathrm{dir}$, but with a modified PDF $\tilde{p}^{j}_{N_j}$. This PDF does not use explicitly all hit times but only the first one $t_1$ and the total number of hits $N_j$. It can be simply derived from the above and one obtains
\[
\tilde{p}^{j}_{N_j}=N_j p_1^j \left( \int_{t_1}^{\infty} \drm t\, p_1^j \right)^{N_j-1}=
N_j p_1^j \left(1- P_1^j \right)^{N_j-1}
\]
where $P_1^j$ is the cumulative of $p_1^j$.

Performance of the reconstruction in IceCube will be explicitly illustrated in a subsequent Section~\ref{sct:ptsrc2}.

\vskip3mm
\paragraph{Generalization}
The log likelihood expression can be formulated generically for the case of composite events~\cite{composite-ribordy}, i.e. events whose mixed detector response is due to the juxtaposition of more than one Cherenkov light source. Such a reconstruction is useful to reconstruct an arbitrary event topology and to favor or discard a given event topology hypothesis (by contrasting $\ln{\mathcal{L}}$ of the various hypotheses). The likelihood minimization becomes slightly more difficult due to the increase of parameters and the fact that  $\ln{\mathcal{L}}$ cannot be split into the two directional and energy components, which are entangled. 

Composite topologies will result for instance from
\begin{enumerate} 
\item  Down-going uncorrelated muons (originating from distinct atmospheric showers) track events. We expect a rate of a few per second at 2 km depth in cubic km-scale detectors;
\item Events from muons undergoing catastrophic energy losses leave a signature resulting from the superposition of showers and a track;
\item Neutrinos converting in the detector effective volume (a shower and a track, possibly finite);
\item Exotic channels (stau pairs, micro black hole evaporation) may result in nearly parallel track events~\cite{staus,mubh};
\item PeV tau double bang and lollipop events.
\end{enumerate}

\subsection{Systematic uncertainties}
Besides intrinsic difficulties (e.g. kinematics of the interaction) and statistical limitations (ET flux versus atmospheric neutrino background), various source of systematics uncertainties contribute to the adversity faced by analyses, which result from our limited knowledge of the instrumental and physical input parameters:
\begin{itemize}
\item[-] Bioluminescence in water, optical properties in the ice: dependence of absorption and scattering profile and length w.r.t. the depth and refrozen ice holes;
\item[-] Optical module absolute sensitivities and their exact locations and orientations;
\item[-] Interaction cross sections (rising uncertainties at UHE); muon propagation;
\item[-] Atmospheric background spectral shape and absolute normalization (cosmic ray spectra, composition and subsequent muon and neutrino production). 
\end{itemize}
These uncertainties cannot be reduced by increasing statistics, at least not in a straightforward manner: some are in principle reducible {\bf indirectly} with increasing statistics, such as the ice properties, the difficulty residing in the decoupling of entangled uncertainties.

\section{Neutrino telescopes: past and present}
Greisen posits the concept of neutrino telescopes in 1960~\cite{ref:inception1},
{\it "...we propose a large Cherenkov counter, about 15~m in diameter, ..."
"...about 3000 tons of inexpensive liquid. ..."
"... from the Crab nebula the neutrino energy emission is expected to be three times the rate of energy dissipation by the electrons, ..."}
and concludes {\it "Fanciful though this proposal seems, we suspect that within the next decade, cosmic ray neutrino detection will become one of the tools of both physics and astronomy."}

Reines discusses simultaneously the arguments, which remain valid nowadays~\cite{ref:inception2}: 
{\it "... [cosmic neutrinos] propagate essentially unchanged in direction and energy ... and so carry information which may be unique in character. For example, cosmic neutrinos can reach us from other galaxies whereas the charged cosmic ray primaries reaching us may be largely constrained by the galactic magnetic field ...."} and notice, due to the lack of understanding of the origin and propagation of the charged cosmic rays:
{\it "... the cosmic neutrino flux can not  be usefully predicted."}
and add {\it "The situation is somewhat simpler in the case of cosmic-ray neutrinos\footnote{note: atmospheric neutrinos}: they are both more predictable and of less intrinsic interest."}
We appreciate now the irony of this statement.

\subsection{First generation}
In 1976, the precursor DUMAND  (Deep Underwater Muon And Neutrino Detector) project was initiated~\cite{dumand}. The early sketch had the ambition of a cubic kilometer telescope, but was downscaled several times. A prototype string (DUMAND-I) was successfully operated from a ship off Hawaii in 1987 at a 4 km depth.  In 1993, elements for a 9-string detector array DUMAND-II consisting of 216 photomultiplier tubes and corresponding front end electronics encased in pressure spheres were essentially ready. Deployment of the infrastructure and a string began in December 1993. Unfortunately, the leaking of an electrical penetrator and the subsequent failure of the string controller electronics  lead to the interruption of the observations after 10 hours of operation. The project was eventually canceled in 1994 due to a lack of funding.

Parallel to the DUMAND project,
the Baikal neutrino telescope effort started in the early 80's~\cite{baikal} and was the first to observe high energy neutrinos~\cite{baikal-nu} in a four string configuration in 1996. The detector located 1.1 km underwater at the Baikal Lake in Russian was expanded into the NT200 configuration in 1998 and then into NT200+ configuration in 2007 with the addition of three sparsely instrumented outer strings~\cite{baikal-nt200plus}. Maintenance and deployment take place from the frozen surface of the lake in winter.

The realization of the AMANDA (Antarctic Muon And Neutrino Detector Array) project was successful after overcoming a certain number of difficulties. In contrast to DUMAND and Baikal, AMANDA was based on an alternative concept, using the Antarctic ice cap rather than water as the deep underground detector medium and target, escaping some of the limiting  water backgrounds (bioluminescence and seawater $^{40}$K decay).
A first string equipped with four optical modules (pressure sphere identical to DUMAND) was deployed in 1992 and successfully operated at 800~m depth. In 1994, four additional strings with 20 optical modules each were installed (AMANDA-A), including optical fibers for laser calibration. It quickly became clear that the characteristics of the ice at this depth were not appropriate because of the contamination with air bubble scattering centers: the Cherenkov light effective scattering length lying at values about 20~cm, i.e. light was heavily delayed when propagating, prevented any reliable reconstruction of the incidence muon directions and therefore efficient searches for neutrinos to be conducted~\cite{Askebjer:1994yn}. This experience nevertheless confirmed the viability of the concept.

\begin{figure}
\centering
\includegraphics*[scale=.4]{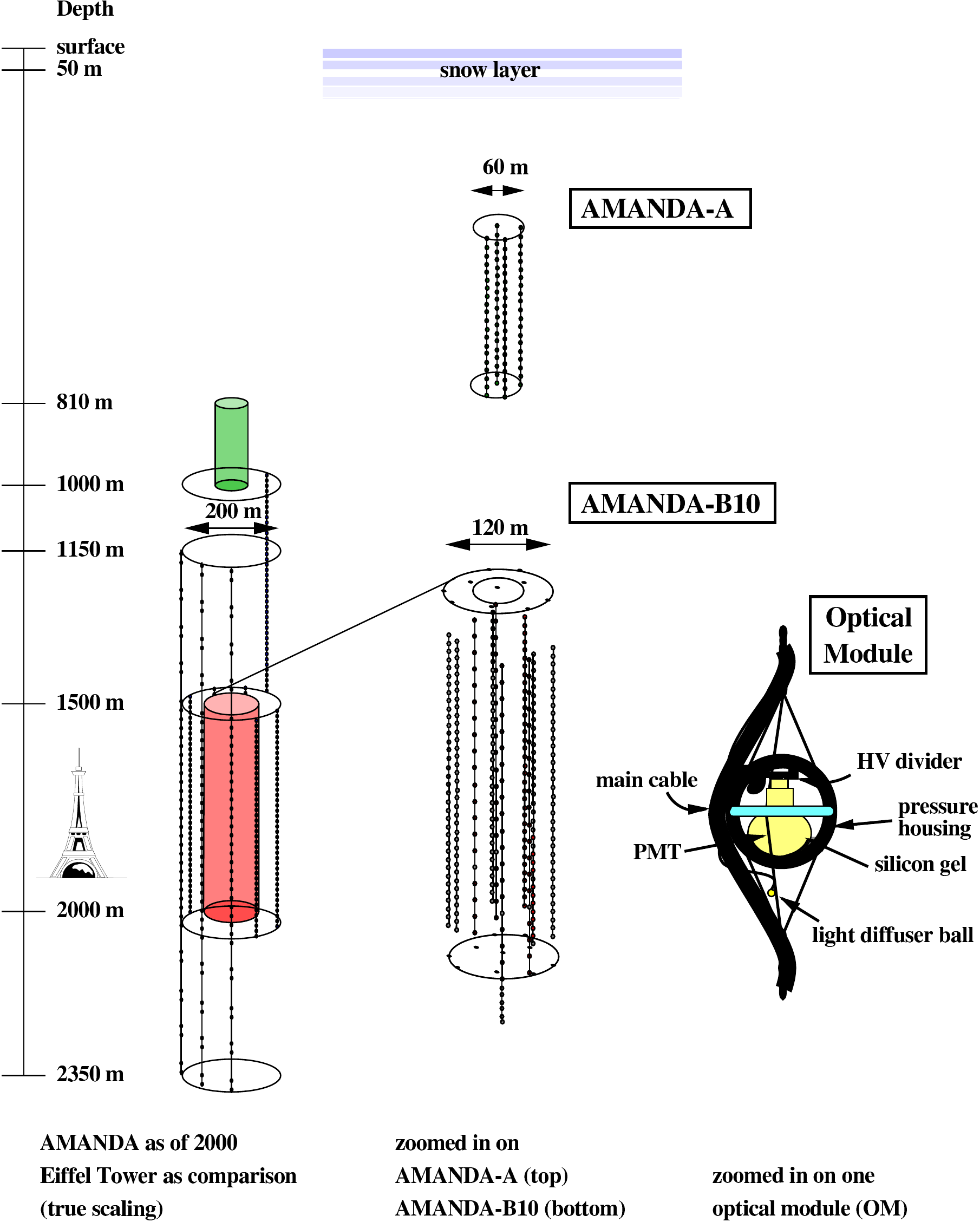}
\caption{The AMANDA neutrino telescope.}
\label{fig:amandaII}
\end{figure}

Ice cores showed that bubble size and density were decreasing drastically  deeper in the ice. The prospects of a deep detector would be further improved considering the additional advantage of a reduction of the atmospheric muon background. In 1997, the AMANDA-B10 detector was deployed and operating~\cite{Halzen:1997za,Andres:1999hm} and it observed atmospheric neutrinos~\cite{Andres:2001ty}. In 2000, an upgrade phase started and nine more strings were deployed (AMANDA-II). Strings are located at depths between 1500 and 2000~m.
Most notably, AMANDA-II has produced very stringent limits on diffuse neutrino fluxes~\cite{Ackermann:2007km,Achterberg:2007qp} and neutrino flux from point sources~\cite{Abbasi:2008ih} and accurately measured the atmospheric neutrino spectrum~\cite{Abbasi:2010qv,Abbasi:2009nfa}. The absence of evidence for ET neutrinos has provided constraints for the future of high energy neutrino astronomy, setting detector scales to cubic kilometer instrumented volumes.

A recent review~\cite{Katz:2011ke} presents the initial pioneering neutrino astronomy phase in greater detail.

\subsection{Present generation}
There are two principal neutrino telescopes in operation: Antares in the Mediterranean and the IceCube neutrino observatory, which signals the emergence of a new class of gigantic detectors dedicated to the observation of the high energy (HE) neutrino sky. The vast research program is found in~\cite{AdrianMartinez:2011ax,IceCube:2011ad,IceCube:2011af,IceCube:2011ag,IceCube:2011ae,IceCube:2011ac,IceCube:2011ah}.

\paragraph{IceCube}
The IceCube observatory was completed in December 2010, after a very successful deployment over the past five years~\cite{icecube,icecube2}.
The in-ice array is equipped with 86 strings, a nominal string spacing of 125~m
Each string is equipped with 60 optical modules, enclosing a photomultiplier and digitization / communication / time stamping boards~\cite{Abbasi:2010vc,Ackermann:2006gp}. The design, construction and deployment of IceCube has largely profited of the gained expertise from the AMANDA enterprise at the South Pole~\cite{Andres:1999hm,Andres:2001ty}, which operated between 1997 and 2009 and provided a proof of principle~\cite{Ahrens:2002gq}.

In the center of the in-ice array, 8 strings more densely instrumented form together with the neighboring strings a dense 20 Megaton inner core, enhancing the IceCube detection capabilities toward lower energies (the data are enriched with lower energy neutrinos) and potentially enabling IceCube to explore the Southern neutrino sky. The design is meeting performance expectations \cite{Achterberg:2006md} and IceCube is taking data at a rate of  $>$2 kHz, dominated by CR muons.

On the ground surface, on top of in-ice IceCube array, the IceTop air shower array instruments a square km. IceTop is dedicated to composition studies of the CR spectrum around the knee and above \cite{Ahrens:2004nn}. IceTop enables the detection of air showers at energies above $\approx$1 PeV. As a part of IceCube, it also plays an important role of facilitating calibration of events triggered by the in-ice array \cite{Ahrens:2004dg} and background rejection of very high energy events.

At the deployment depths, the secondary muon intensity originating from cosmic air showers in the atmosphere is reduced by a large factor. This flux remains the major experimental background for these detectors. The soft atmospheric neutrino background constitutes an irreducible background for ET neutrino searches, which can only be compensated by expected harder signal energy spectra.

IceCube's main goals are the unambiguous identification of the first galactic and extra-galactic cosmic ray (CR) accelerators with the detection of HE neutrinos from point sources and the divulgation of the nature of dark matter (DM) through the observation of a secondary neutrino flux from annihilating DM in our galaxy.

\begin{figure}[ht]
\centering
\includegraphics*[scale=.55]{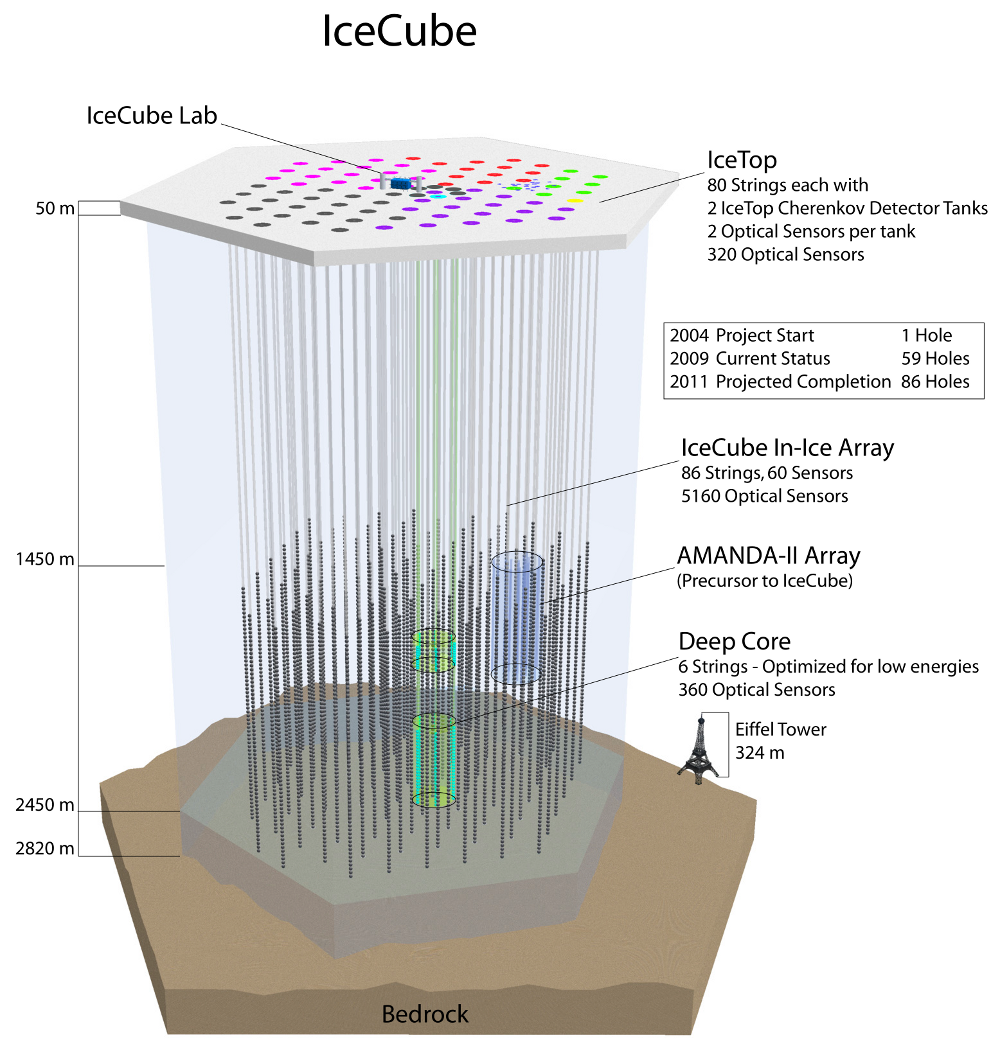}~~~~~~
\includegraphics*[scale=.3,angle=90]{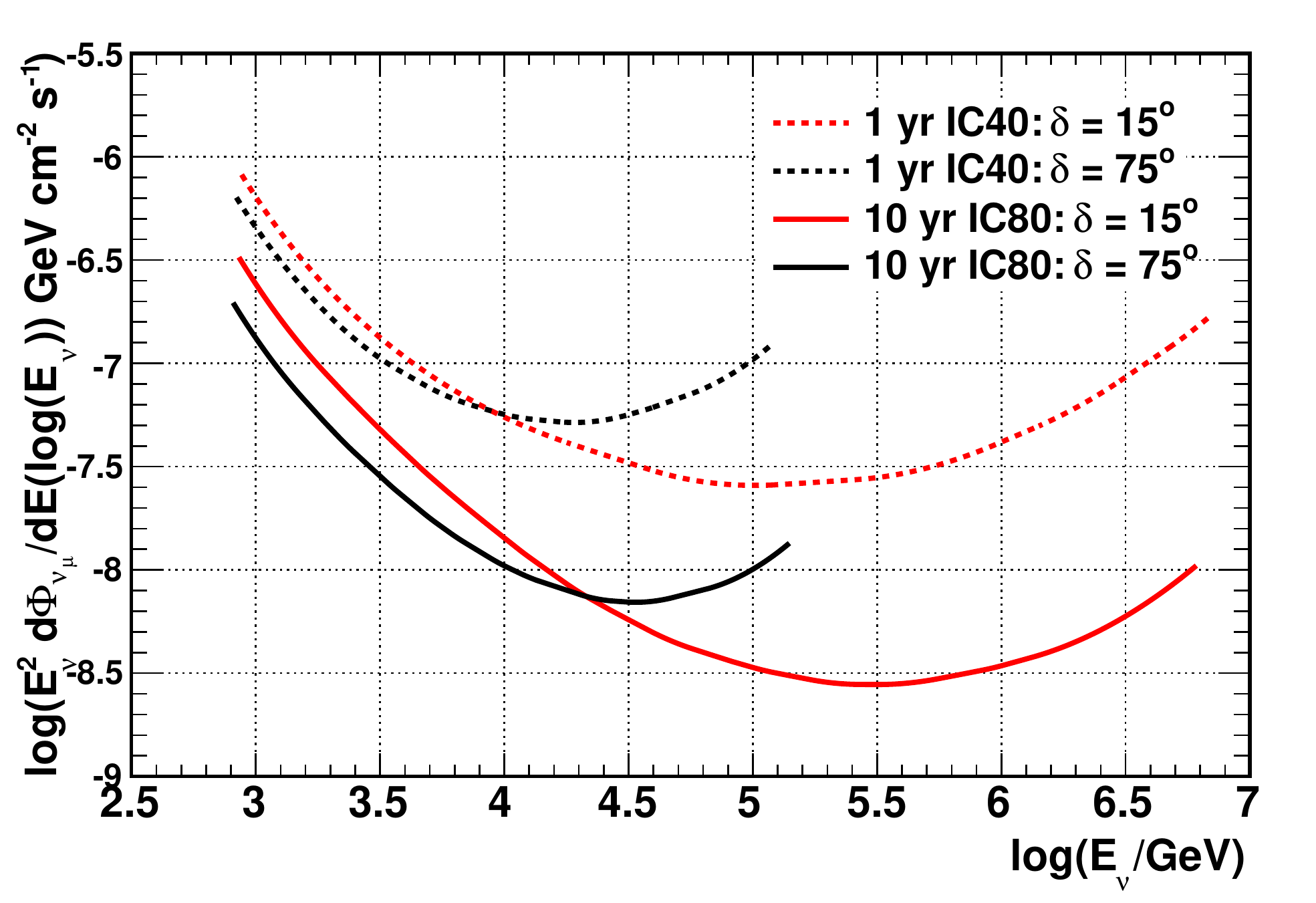}
\caption{Left: IceCube. Right: comparison of the IceCube point source flux discovery potential for one year with IceCube in its 40 string configuration and after ten years with the completed array for different source declinations.}
\label{fig:i3-aeff}
\end{figure}

The rate of increase of the IceCube discovery potential has been greatest during the deployment phase and will continue to be high until about 3 years of data have been analyzed. A noticeable that a jump by an order of magnitude of the discovery potential will occur within the next 10 years (for a 20-fold increase in the statistics), see Fig.~\ref{fig:i3-aeff}, right, compared to current results~\cite{Abbasi:2010rd}. In the case of point source discovery during this period, an era of precision for HE neutrino astroparticle physics will be initiated and the prospects will include the characterization of the detected neutrino fluxes (i.e. the energy spectra and their time-dependence).

On site, prospecting activities for the detection of UHE neutrinos are taking place as well, relying on the alternate radio Cherenkov and acoustic signatures accompanying UHE neutrino interactions, see Section~\ref{sct:prospects}.

\paragraph{Antares}
Antares operates in the Mediterranean See since May 2008 at 2.5 km depth and consists of 900 PMTs distributed on 25 storeys and 12 lines (3 PMT per storey), placed 60-80 m apart, thus instrumenting a volume of 0.025 km$^3$. Not located at a Pole, the field of view is continually changing and exposure is not uniform but complementary to IceCube. It consists of 13 strings half a kilometer long, deployed at a depth of 2.5 km. An excellent muon angular resolution of 0.1$^\circ$ can be achieved due to the low scattering of Cherenkov light. In this non sterile environment, one has nevertheless to cope with currents, organic lightning, dirt deposit on the optical modules and with potassium decay, leading to 100 kHz PMT noise rate, necessitating stages to be equipped with 3 modules operated in coincidence.

\section{Supernova neutrino detection methodology}
The core collapse supernova energy release is mainly directed into the neutrino component.  Estimates posit the frequency of these events in the Milky Way of about $2\pm1$ per century, so that expectations are optimistically 25\% probability of such an event in the next decade~\cite{Li:2010kd}. Such an event would be of great significance, not only constraining the dynamics of the core collapse itself but potentially granting access to neutrino hierarchy (remember that the sign of difference of mass eigenstates is accessible through matter oscillations and scales with $\theta_{13}$).

The main phases of 10--100 MeV neutrino emission are successively
\begin{enumerate}
\item $\mathcal{O}(10\,\text{ms})$ deleptonization phase, a burst of electron neutrinos;
\item Few 100's ms accretion phase;
\item Cooling phase, with roughly all flavor energy equipartition.
\end{enumerate}
Far more detail on the physics of supernova can be found in the dedicated lectures by G. Raffelt in these school proceedings.

\subsection{Neutrino telescope as supernova detector}
Neutrino telescopes operating in the quiet ice medium can be utilized as instruments for supernova detection, due to the very large neutrino flux from galactic supernova at earth and moderate optical module noise, of order $r=0.58$~kHz, this even while the cross section are very small and low photon yield signature of individual event is inappropriate for these types of instruments.

The traditional detection principle is based on the sudden increase of the detector hit noise rate, as originally proposed~\cite{pryor,Halzen:1995ex,Halzen:1994xe} and IceCube can monitor the whole Milky Way~\cite{Abbasi:2011ss}.
The main detection channel is the interaction of anti-electron neutrinos with hydrogen protons, $\bar\nu_{\rm{e}} {\rm{p}} \rightarrow {\rm{e}}^{+}{\rm{n}}$. The contribution of other channels is suppressed due to the large oxygen $^{16}{\rm{O}}$ binding energy, the rare presence of oxygen and hydrogen isotopes and the small electronic cross sections. A significant fraction of hits (about 10\%) arise from neutral current cross section with oxygen.

Mediterranean telescopes cannot be effectively used the same way, due to the large $^{40}{\rm{K}}$ radioactivity and microscopic bioluminescence (both components accounting for about  100 kHz optical module noise rate in quiet conditions).

The brief and powerful electron neutrino neutronization emission is of great importance given its connection to astrophysics and particle physics and can be studied via electron recoil in elastic collisions and to some extent through oxygen charged current reaction. The burst luminosity  also moderately depends of the progenitor mass, granting potentially access to the distance of an obscured core collapse supernova located at 10 kpc with 5\% precision~\cite{snKachelriess}.

Let's consider a benchmark supernova at $d=10$~kpc releasing $E_{\rm{SN}}=2\times 10^{59}$ MeV in neutrinos
 within about $T=3$ seconds (a nearly optimal integration time in a simple analysis to assess the detection potential in SF model~\cite{garching}), we can calculate the time-averaged $\bar\nu_{\rm{e}}$ flux, \[\bar\Phi_{\bar\nu_{\rm{e}}}(d=10\,{\rm{kpc}})=\frac{1}{6}\frac{1}{T}\frac{E_{\rm{SN}}}{\langle E_{\bar\nu_{\rm{e}}}\rangle}\frac{1}{4\pi d^2} \approx10^{11}\,\cm^{-2}\,\second^{-1},\]
where factor $1/6$ reflects flavor equipartition.

In the energy range of interest, the interaction cross section is quadratically growing with energy, about $\sigma_{\bar\nu_{\rm{e}}{\rm{p}}}\approx 10^{-41}\,\cm^{2}$ at 12.5 MeV.
We obtain the interaction rate in a gigaton detector volume $V$ with density $\rho$, 
\[
\rho V \bar\Phi_{\bar\nu_{\rm{e}}} \sigma_{\bar\nu_{\rm{e}}{\rm{p}}} N_a/9 \approx 0.7\cdot 10^8\,\second^{-1}
\]

Note that it would be more correct to use in this case a cross section weighted on the differential neutrino flux,
\[
\int\drm E_\nu \int\drm E_{\rm{e}}
\frac{\drm\sigma_{\bar\nu_{\rm{e}}\proton}}{{\rm{d}}E_{\rm{e}}}(E_\nu,E_{\rm{e}})\,
\frac{\drm\bar\Phi_{\bar\nu_\electron}}{\drm E_\nu}(E_\nu,\,d)\,
\]
in place $ \bar\Phi_{\bar\nu_{\rm{e}}} \sigma_{\bar\nu_{\rm{e}}{\rm{p}}}$, but we drop this for simplicity in this back of the envelope calculation. It would also turn out that the average positron energy is significantly higher than the average neutrino energy as shown in Fig.~\ref{fig:SNenergy}.

Positrons from $12.5$ MeV neutrinos will roughly have energy $E_{\rm{e}} = E_\nu - (\Delta m_{pn} + m_e)$, where $\Delta m_{pn} + m_e \approx 1.8\,\rm{MeV}$, i.e. emit about $N(E_\electron)=2000$ Cherenkov photons between 300 and 650 nm along  a 5 cm track. Due to multiple scattering, the Cherenkov photon emission will be quite uniform over $\pi$ sr.

The optical module in IceCube have a diameter of about $2R=30\,\cm$ and a photo-detection efficiency of about $\epsilon_{\rm{PMT}}=7$\% (photo-multiplier tube quantum efficiency and glass), once integrated between 300 and 650 nm for a Cherenkov spectrum. It is simple to estimate the effective volume of the IceCube detector composed of about $N_{\rm{OM}}=5000$ modules for the detection of uniformly interacting neutrinos  in the detector volume,
\[ V^k_{\rm{eff}}(E_{\rm{e}}) =  2\pi\,N_{\rm{OM}}\, \int_{R}^{\infty} r^2 {\mathrm{d}}r
\int_{-1}^{1} \drm\cos{\theta}\, f(\ge k,\,r,\,\theta,E_{\rm{e}}),\label{snEffV}
\]
where $f(\ge k,\,r,\,\theta)$ is the probability density of recording at least $k$ hits from a positron at distance $r$ from an optical module and at an angle $\theta$ (optical modules are not uniformly sensitive, so we define $\theta$ as the angle between the optical module axis and the positron with the optical module at the origin of the coordinate system).

For $k=1$, we have
\[f(\ge 1,\,r,\,\theta,E_{\rm{e}})\approx f(1,\,r,\,\theta,E_{\rm{e}})\approx \epsilon_{\rm{PMT}}N(E_\electron)\,\epsilon(r,\theta)\,{\rm{e}}^{-r/\lambda_{\rm{abs}}}\]
where $\epsilon(r,\theta)$ is the fraction of the $4\pi\,$sr solid angle occupied by the sensitive surface of the module, viewed under angle $\theta$ and at distance $r$ and $\lambda_{\rm{abs}}$ is the medium absorption length of the Cherenkov photons.
$V^1_{\rm{eff}}$ is slightly above 2 megatons, i.e. about 0.2\% of the instrumented volume, or a hit rate of 200 kHz, well above the nominal standard deviation of the detector noise rate $\sqrt{N_{\rm{OM}}(r/{\rm{Hz}})}\,{\rm{Hz}}\approx 1.7\,{\rm{kHz}}$. Integrated over the optimal time window $T$, we may therefore expect a 5 standard deviation excess up to distances of about 65 kpc.
Detailed analysis presented in~\cite{Abbasi:2011ss} confirms this picture.

The traditional method is therefore very effective for the detection of galactic supernova and IceCube complements other operating neutrino detectors, e.g. Super-K (few thousands detected neutrinos for this benchmark supernova), useful when an instrument is in maintenance, to cross calibrate, triangulate the SN incidence direction (IceCube resolution is 2 ms) or to study earth matter effect. 

Assuming an emission spectrum,  this method enables the extraction of information related to the supernova luminosity as well, once the supernova distance is known.
However, the neutrino flux and average energy cannot be disentangled. 
Moreover, in the case the supernova escaped optical detection, the luminosity will remain unknown.
It is therefore interesting to consider alternative detection channels.

\subsection{SN detection based on coincident hit rate}
%
Coincident hits are hits at a single or neighboring optical modules, occurring within a very short time window, of about $\le 100\,{\rm{ns}}$ and originating from a single positron. The ratio of coincident to single hit rates increases with the average positron energy: while the single hit probability is rising linearly with the positron energy (i.e. the positron track length and number of emitted Cherenkov photons), the probability for two correlated hits is rising with the square of the positron energy.

Therefore this ratio constitutes a new observable for the measurement of the average energy of the supernova neutrinos. Combining it with the single hit rate, it enables to disentangle neutrino flux and average energy, in contrast to the traditional supernova detection method.

\begin{figure}
\centering
\includegraphics*[scale=.9]{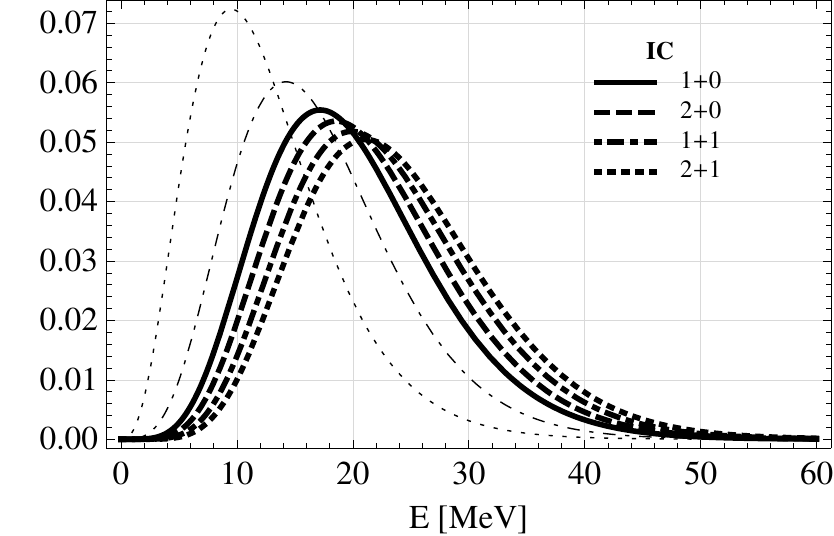}
\caption{From left to right. SF model neutrino emission spectrum (thin, dotted) and positron
    spectra: produced (thin,
    dot-dashed), producing single hit (solid), double hit (dashed),
    $1+1$ (dot-dashed) and $2+1$ (dotted) signatures for IC.}
\label{fig:SNenergy}
\end{figure}
Let's discuss here only the case of two coincident hits at a single module for simplified treatment and {\it Sf} neutrino emission spectral shape model:  the average energy of the positrons, about $17\,\mev$, is significantly larger than the neutrino energy,  about $\approx12.5\,\mev$, due to the rising neutrino interaction cross section with energy. Average energies of the positrons resulting in a single and double hits are respectively about 20 and 22 MeV, illustrated in Fig.~\ref{fig:SNenergy}. Using these numbers and eq.~(\ref{snEffV}), the ratio of single to double hits can be calculated and one obtains about 1\%. This ratio sensitively depends on the average energy and few percent accuracy would be obtained for the benchmark supernova. 

\subsection{Backgrounds and other difficulties in SN analyses}
The analysis of supernova neutrinos has to cope with the atmospheric muon background, which contribute about 15 Hz to each optical module hit rate in average. This background is also at the origin of correlated hits in neighboring optical modules and can be reduced by removing external detector layers (atmospheric muons dying in the superficial detector layers and producing hits insufficient to fulfill the event trigger condition). Seasonal variations of module hit rates, due to seasonal change of the temperature and pressure in the atmosphere, also affect the analysis and outcome in term of significance of an observation and must be carefully accounted for. 
Sensor correlated hits at second and milli-second time scales, due respectively to photomultiplier afterpulse and radioactive decay chains may eventually also reduce
the coincident hit method potential.

Calculation details, rejection of correlated hits  from muon background, results from simulation  and prospects with future IceCube extensions are found in~\cite{Demiroers:2011am}.

\section{Point source search methodology}

\subsection{The multi-messenger approach}
The existence of a correlation between the energy released by an astronomical source in the form of electromagnetic (EM) radiation, CRs and neutrino flux is at the origin of the multi-messenger (MM) paradigm, supported by convincing phenomenological arguments and numerous models of source activity.

The most basic expression of analyses performed in the MM context is the search for a significant excess of neutrinos (over the atmospheric neutrino background) from the direction of point source candidates selected on the basis of their MWL spectra along with the appropriate phenomenological arguments. This is diametrically opposed to a random search for neutrino point sources (e.g. a whole sky search), which has no direct connection to phenomenology (with IceCube, it has been demonstrated that the analysis method is robust with a change of the assumed spectral index as we will see later). The MM approach augments the significance of observation, however, this is true only if the actual neutrino sources are among the selected targets: the approach relies on the validity of phenomenological arguments; moreover, the existence of neutrino sources not emitting in the electromagnetic spectrum cannot be excluded.

\subsection{Point source search strategies}
Point source searches are analyses looking for a statistical excess originating in narrow sky regions. Best current sensitivities are rather uniform w.r.t. the declination and notably below the intrinsic TeV radiation strength of e.g. Mrk 501 during the 1997 flaring state (assuming $\gamma/\nu \approx 1$), These searches are optimized by taking advantage of the experimentally observed off-source detector response, which defines the atmospheric neutrino background {\it i.e.}, at first order, sources of systematic errors can be neglected.

Steady point source searches, on the whole sky or based on catalogs, are conducted within the HE neutrino telescope community~\cite{Abbasi:2010rd,Abbasi:2009cv,Abbasi:2009iv,Abbasi:2008ih,Dornic:2011wf}. Concurrent searches, based on MWL spectra, introduce an optimization based on the spectral characteristics and/or the time dependence of the photon emission in a certain energy band reflected in the neutrino emssion, for instance for GRB's, periodic galactic systems, flaring objects such as AGN's~\cite{IceCube:2011ai,Abbasi:2011ke,Abbasi:2011qc,Abbasi:2009ig,Abbasi:2009kq,AdrianMartinez:2011jz}.

\subsection{Whole sky steady point source search}\label{sct:ptsrc2}
We discuss now the generic search methodology for localized excess in the sky in greater detail. We start with the very basic approach, so called "binned" search and restrict to the very specific South Pole location for simplicity, as the background density is a function of declination $\delta$ only:  the zenith angle $\theta$ of the incoming neutrino direction is related to $\delta$ independent of time and the right ascension $\alpha$ drops assuming a cylindrical detector symmetry and steady emission. 

Consider a circular or square bin of area $a$ around a source located at $(\delta,\alpha)$. The bin extension is chosen in relation to the point spread function $\Theta_{\nu_l l}(E)$ for the entire neutrino sample at this declination.
Given $N$ neutrino candidate events in the sample in the $\delta$ declination band of area $A$, the bin background expectation is \[\mu_{\rm{b}}= N p= N(a/A).\]
The significance of an observation of  $n_{\rm{obs}}$ events in the search bin in the absence of signal is obtained using the cumulative distribution function (CDF) $I_B$ of the binomial probability $P_B(n_{\rm{obs}};\,N,\,\mu_{\rm{b}})$.
In point source searches, the significance of an observation can be set without relying on the Monte Carlo. Obviously this is not the case when it comes in setting a flux or flux limit: a detector response model is necessary. In practice, the event candidate selection is made with the help of the Monte Carlo, in order to adjust the analysis to a specific assumption on the source spectrum (usually taken as a power law spectrum with $\gamma=2$ extending to very high energies, eventually dominating the background of atmospheric neutrinos)

In order to avoid any bias in the selection of the neutrino candidates, the events selection optimization is performed "off-source", using randomized right ascension $\alpha$, which is, still restricting the approach to the South Pole location, equivalent to scrambling the recorded event timestamps.

\paragraph{Concept of upper limit and sensitivity and discovery flux}

Once one disposes of an actual measurement, assumed to be below the discovery threshold arbitrarily set to something like a 5$\sigma$ significance of non zero observation, the upper limit at some confidence level (C.L.) can be set (usually something like 90\%).
The meaning of the 90\% C.L. upper limit is the following: for a given experimental outcome $n_{\rm{obs}}$, the {\bf true unknown} signal lies below the 90\% flux upper limit $\Phi_{90}$ (or $\mu_{90}$ in terms of number of events) in 90\% of the experiments~\cite{Hill:2002nv}. The construction of upper limits presented in the literature often relies on the Neyman~\cite{neyman} or Feldman-Cousin~\cite{feldman-cousins,cousins} prescriptions. The conversion of $\mu_{90}$ into $\Phi_{90}$ necessitates the Monte Carlo as already stated and must assume a spectrum.

Before the actual measurement takes place (i.e. the experimental data analysis), con\-cur\-rent analyses must be tuned and performance compared. A useful concept is the 90\% average upper limit or {\bf sensitivity}. As the name indicates, the sensitivity is the upper limit averaged for all possible outcomes $n_{\rm{obs}}$ of an ensemble of experiments {\it in the absence of true signal},
\[
\bar\mu_{90}(\mu_{\rm{b}}) = \sum_{n_{\rm{obs}}=0}^{\infty}  \mu_{90}(n_{\rm{obs}},\,\mu_{\rm{b}}) \,\frac{\mu_{\rm{b}}^{n_{\rm{obs}}}}{n_{\rm{obs}} !}\, {\rm{e}}^{-\mu_{\rm{b}}}
\]

The power for discovery of an experiment after a certain operating time is another useful analysis outcome. The literature sometimes mentions  the {\bf $5\sigma$ discovery flux in 50\% of the experiments}. In the simple case of a binned search, this can be built easily: the binomial PDF defines the minimum $n_{\rm{obs}, 5\sigma}$ for a detection at the $5\sigma$ C.L. given $\mu_{\rm{b}}$. The discovery flux is set by finding the right amount of signal $\mu_{\rm{s}}$  to add to the binomial PDF
\[ P_B(\mu_{\rm{b}})
\rightarrow P_B(\mu_{\rm{b}}+\mu_{\rm{s}})
\]
necessary to ensure an experimental outcome $n_{\rm{obs}} \ge n_{\rm{obs}, 5\sigma}$ in 50\% of the experiments.

This binned search is appropriate to obtain the results analytically, i.e. building the test statistics (TS) $I_B$ and $\bar I_B$ and $\mu_{\rm{s}}$ can be extracted analytically
\[\mu_{\rm{s}} \,|\, \{I_B( \mu_{\rm{b}}+\mu_{\rm{s}},n_{\rm{obs}, 5\sigma})={\scriptsize\frac{1}{2}}\},\]
where \[n_{\rm{obs}, 5\sigma}=n_{\rm{obs}} \,|\,  \{\bar I_B(\mu_{\rm{b}}, n_{\rm{obs}})=\frac{1}{2} \text{erfc}(5/\sqrt{2})\}\,\text{ ~~~~~~~ (@}\,5\sigma\,\text{C.L.)}.\]

In practice, and in order to introduce the concept once we will consider the more powerful "unbinned" search methodology (which, as a drawback, relies heavily on the Monte Carlo), the construction of the TS illustrated Fig.~\ref{fig:TS} proceeds according to the following steps: 
\begin{figure}[ht]
\centering
\includegraphics*[scale=.35]{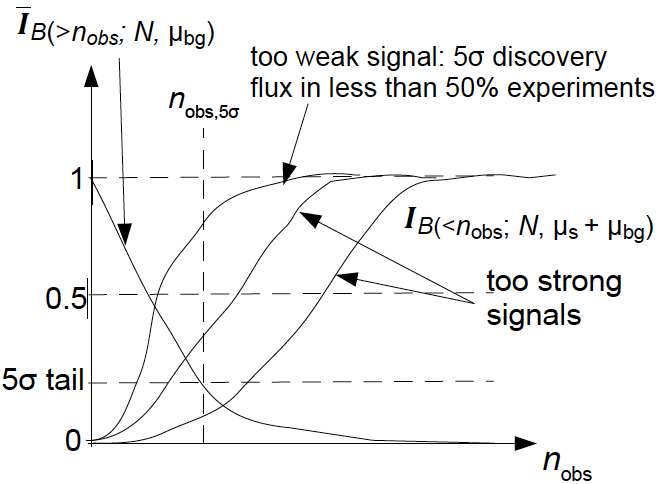}
\caption{Illustration of the building of the test statistics in the simplified case of a binned search.}
\label{fig:TS}
\end{figure}
\begin{enumerate} 
\item By means of experimental and Monte Carlo data, the TS is generated by scrambling the sky maps of experimental data million of times including some variable amount of Monte Carlo signal according to the assumptions of source location and spectrum;
\item For each map, the TS distribution is filled with $n_{\rm{obs}}$ in the search bin around $(\delta, \alpha)$.
\end{enumerate}

The discovery flux is found by comparing the null signal CDF with the set of CDF's with non vanishing signal. The discovery flux leads to $n_{\rm{obs}} \ge n_{\rm{obs}, 5\sigma}$ in 50\% of the experiments, where $n_{\rm{obs}, 5\sigma}$ is the number of events in the absence of signal that would be attained by a fraction $1/1.7\cdot10^6$ of the experiments.

\paragraph{Generalization: "unbinned" search}
We discuss now the straightforward generalization of the methodology to "unbinned" point source searches~\cite{Braun}. The  additional power originates in the non binary underlying methodology: in the binned method, no maximum profit was taken of the shapes of the angular PSF and energy PDF, i.e. we only minimally exploited our knowledge of the detector and physics (kinematics, muon energy at the detector). We now apply a maximum likelihood fit to the data to determine the relative signal contribution.

We assume a data sample, which  consists of $N$ neutrino candidates, as a mixture of two components, signal and background. The probability density of the $i$th event measured at location $(\delta_i,\alpha_i)$ with energy $E_i$, assuming a signal source with spectral index $\gamma$ located at $(\delta, \alpha)$  is 
\[ \frac{n_{\rm{s}}}{N}{\cal S}_i +(1-\frac{n_{\rm{s}}}{N}){\cal B}_i, \label{eq:unbinnedPDF}\]
where the signal and background PDF's are
\begin{eqnarray}
{\cal S}_i &=& {\cal N}_{\rm{s}}(\{\delta_i, \alpha_i\}\,|\, \{\delta, \alpha\},\,\sigma_i) \times  { \cal E}(E_i\,|\,\gamma, \delta_i)
\\
{\cal B}_i &=& {\cal N}_{\rm{atm}}(\delta_i) \times  {\cal E}(E_i\,|\,\text{atm}, \delta_i)
\end{eqnarray}
Here ${\cal N}_{\rm{s}}$ is the probability that an event originates from the source and modeled as a 2-D gaussian of width $\sigma_i$. $\sigma_i$ is the angular resolution obtained for each event individually from the error on the parameters from the  maximum likelihood reconstruction fit\footnote{More precisely, $\sigma_i$ is obtained from a faster algorithm (not from the error processors available in the MINUIT package), the result from a paraboloid fit of the likelihood landscape space of the reconstructed incoming direction of the event $(\theta,\phi)$ is an estimator of the angular uncertainty.}; ${\cal N}_{\rm{s}}$ is the angular PDF for the atmospheric background and is flat w.r.t. $\alpha$ at the South Pole; ${\cal E}$ describes the probability of the reconstructed energy $E_i$.

\begin{figure}[here]
\centering
\includegraphics*[scale=.45]{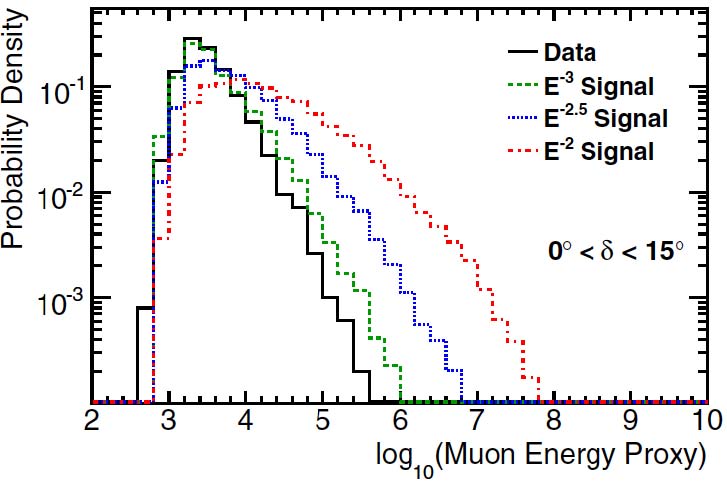}
\includegraphics*[scale=.45]{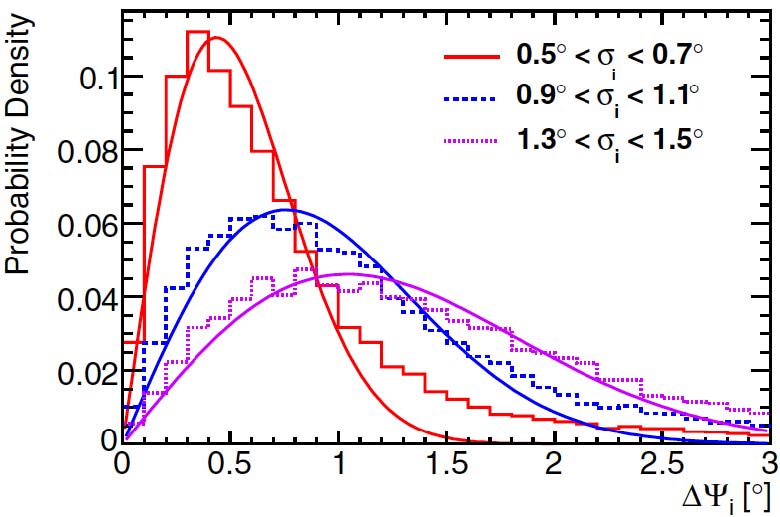}
\caption{Estimators. Figs from~\cite{Abbasi:2010rd}.}
\label{fig:pdf-unbinned}
\end{figure}

Fig.~\ref{fig:pdf-unbinned}, left, shows the PDF of the muon energy estimator for various simulated signal power laws and for experimental data, while Fig.~\ref{fig:pdf-unbinned}, right, shows for simulated data the correlation between $\sigma_i$ and the actual track reconstruction error. 

The parameters $\tilde{n}_{\rm{s}}$ and $\tilde{\gamma}$ are obtained from a maximum likelihood on all events $i\in\{1, N\}$ using the PDF from eq.~(\ref{eq:unbinnedPDF}),
\[ {\cal L}(n_{\rm{s}}, \gamma) = \prod_{i=1}^N \left[ \frac{n_{\rm{s}}}{N}{\cal S}_i +(1-\frac{n_{\rm{s}}}{N}){\cal B}_i, \right].
\]
The TS distribution, built with the millions of scrambled sky maps, is commonly defined against the null hypothesis $n_{\rm{s}}=0$,
\[
\lambda = -2 \ln{\frac{{\cal L}(n_{\rm{s}}=0)}{{\cal L}(\tilde{n}_{\rm{s}}, \tilde{\gamma})}}.
\]

\begin{figure}[here]
\centering
\includegraphics*[scale=.16]{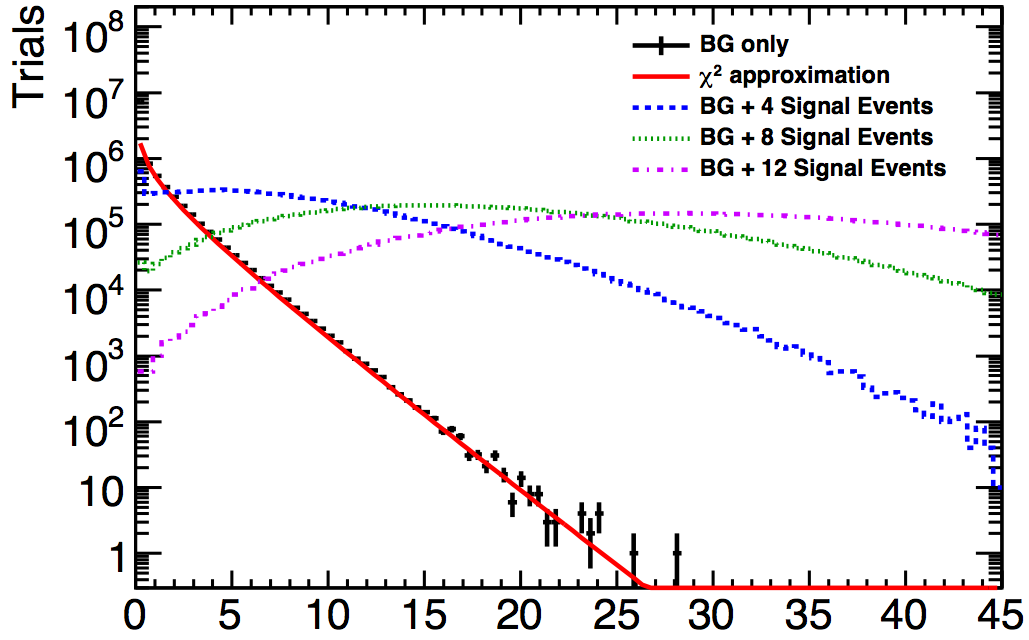}
\includegraphics*[scale=.17]{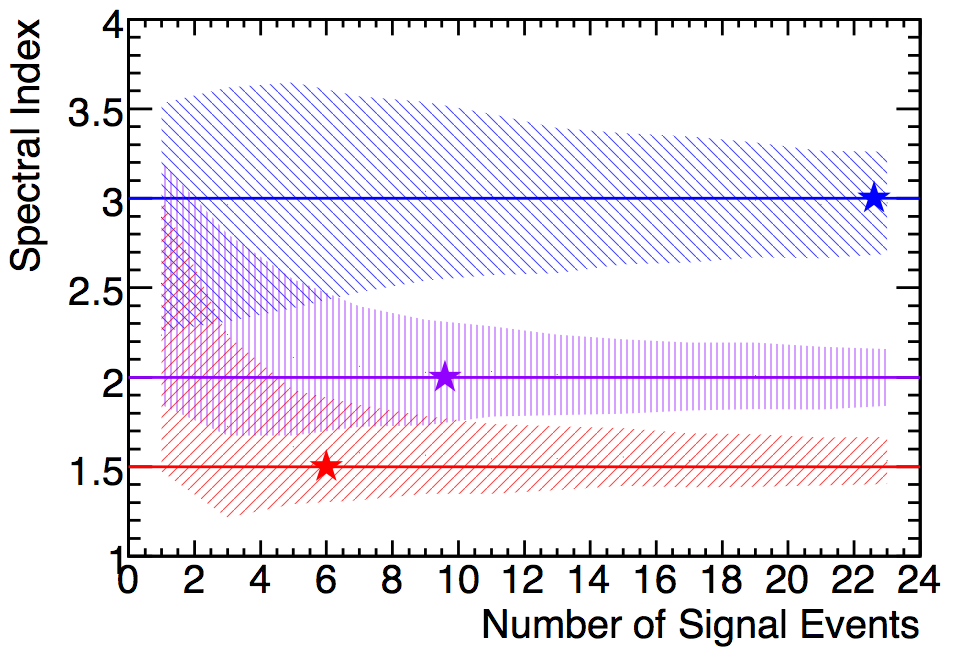}
\caption{Test statistics in the case of an unbinned search and potential of the method (see text). Figs from~\cite{Abbasi:2010rd}.}
\label{fig:TS-unbinned}
\end{figure}

In order to assess the significance of an observation, the following steps must be taken:
\begin{enumerate}
\item Build the TS distribution. Compared to the binned search, it simply consists into the substitution of $n_{\rm{obs}}$ with the two-step process of maximization and calculation of $\lambda$;
\item The final number, the $p$--value associated to $\lambda$ is the fraction of scrambled data sets leading to higher TS values than $\lambda$. $\lambda$ defines the discovery flux, and used in the Feldman-Cousin construct for instance, the sensitivity and upper limit.
\end{enumerate}
These are illustrated on Fig.~\ref{fig:TS-unbinned}, left: we see the TS distributions of pure background and background + various signals; at $\lambda\approx24-25$, the residual fraction of experiments leading to larger $\lambda$ is at the $5\sigma$ level. The signal TS shows that a source with a strength corresponding to a number of selected events between 8 and 12 is necessary to have 50\% of the experiments with  $\lambda>24-25$.

Fig.~\ref{fig:TS-unbinned}, right, demonstrates that this formalism enables the study of the resolution on $\gamma$ and the number of selected source events required for discovery for various spectra, $\gamma=1.5,\, 2,\,3$.

A further complication arises, related to the statistical trial factor. If a search looks for evidence of an excess from a small number of predefined and well separated sources, the trial factor is the number of sources in this catalog. When searching in the whole sky for a point source excess, it is not so well defined: for a given monochromatic source,  the "effective" number of sources would roughly correspond to $4\pi/a(E)$, where $a$ is the subtended area from the PSF width. In practice, the unbinned analysis is repeated on the global sky, filling the TS distribution with the sky location corresponding to the best $p$--value. One obtains the post-trial $p$--value.

\begin{figure}[here]
\centering
\includegraphics*[scale=.16]{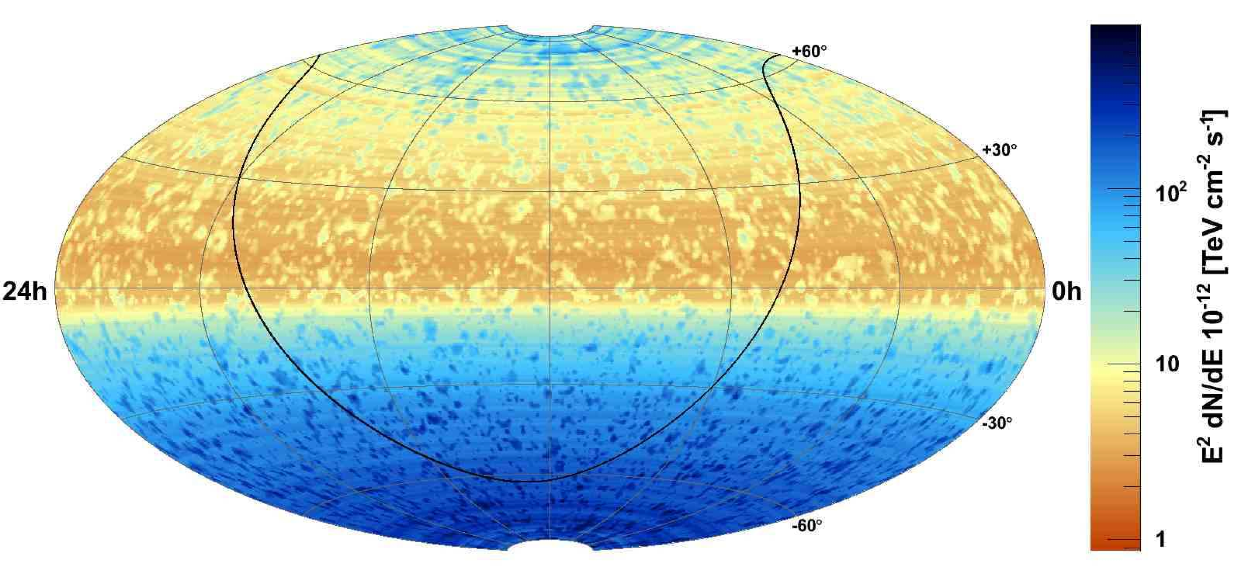}
\caption{Upper limit skymap with IC-40, taken from~\cite{Abbasi:2010rd}.}
\label{fig:skymaps-IC40}
\end{figure}

Fig.~\ref{fig:skymaps-IC40} shows the IC-40 sky in terms of 90\% upper limits. One notices  that they are best for sources at positive declination and are degraded by 10--100 in the Southern sky for obvious reasons.

The unbinned search methodology can be extended in a very straightforward fashion,
\begin{enumerate}
\item For transient and periodic sources (for instance GRB, AGN burst, binary systems), by simply adding the corresponding time-dependent factor in the signal PDF.
\item To stacked searches, which consist in cumulating the signal from similar sources with different weights. E.g. the weight may depend on the source distance or on the source acceptance (different background at different declination at the South Pole, different exposure in ANTARES).
\item To extended point source searches, by replacing the PSF with the PSF convoluted with the source distribution: PSF $\rightarrow$ PSF $\otimes$ source distribution. 
\end{enumerate}

\section{UHE neutrino search with alternative detection techniques}\label{sct:prospects}
Typical predictions for the cosmogenic neutrino flux, i.e. neutrinos from photo-pion production, is of the order $E {\rm{d}}N/{\rm{d}}E \sim 10^{-17} \,\mathrm{s}^{-1}\,\mathrm{cm}^{-2}\mathrm{sr}^{-1}$ at $E=10^{18}$ eV. In ice, this flux translates into one neutrino interaction / km$^3$ every few years. As we mentioned previously, however, the prediction strongly depends on the primary cosmic ray composition among other parameters and the rate could be lower by one or two orders of magnitude.

The characterization of the GZK neutrino flux spectrum, and thus the recovery of the degraded information carried by UHE CR's, would permit the delineation of cosmological source evolution scenarios from source injection spectrum characteristics. To fulfill this goal, however, the GZK event detection rate should be vastly increased. Therefore, a much larger volume should be instrumented with an appropriate technology for the detection of ultra high energy neutrino interactions.

As the applicability of the optical detection technique reaches its limits due to $\lesssim100\,\text{m}$ attenuation length, two alternative detection methods were proposed for operation on large arrays, following signatures first discussed by Askaryan~\cite{aska,learned:1979,bevan:2007}: an interacting neutrino emits a coherent Cherenkov pulse in the range of 0.1-1 GHz \cite{Zas:1991jv} close to the shower axis and a thin thermoacoustic pancake normal to the shower axis~\cite{Dedenko:1997ur}.

The bremsstrahlung and pair production cross sections starts decreasing quickly above 
$\mathcal{O}(\text{PeV})$  strongly affecting the longitudinal profile of the shower above EeV energies, which then consists of disparate subshowers initiated by these UHE daughters. This feature, known as the LPM effect (Landau-Pomeranchuk-Migdal)~\cite{lpm1,lpm2} and fundamentally related to the non point-like nature of the interactions, reduces in particular the prospects of the acoustic detection technique (higher energy threshold).

The radio detection technique exploited by several past, operating or projected instruments observe the direct~\cite{rice,salsa,ara,arianna} or the refracted radio Cherenkov pulse either from the air~\cite{anita} or from space~\cite{forte} following an interaction near the earth's surface, or they observe the signal following an interaction in the lunar regolith \cite{moon_regolith}.

Acoustic sensor arrays deployed in water \cite{acoustic-water} or in ice \cite{spats} rely on the acoustic detection technique~\cite{Nahnhauer:2007zz,Vandenbroucke:2006ax}.

The radio detection technique is traditional at the South Pole, the RICE detector~\cite{rice} operating for a long time has paved the path toward the ARA project~\cite{ara} consisting of a series of antennas in the firn ice layer. The ARA sensitivity to ultra high energy neutrinos would eventually reach the level of unfavorable GZK neutrino flux expectations (pure Fe composition). ARA could be complemented with surface radio stations~\cite{rasta} and, conceivably with acoustic detection devices. 
%

Contrary to salt and water, ice has the unique feature of allowing the detection of three concurrent signatures accompanying a neutrino event, optical Cherenkov light~\cite{ice}, coherent radio Cherenkov and thermoacoustic emissions. The combined detection of two or more of these signatures, leading to a strong background reduction, would unambiguously tag a neutrino event and thus firmly establish the event origin~\cite{Halzen:2003fi,Vandenbroucke:2007pk}.

In ice, the attenuation length of the radio and acoustic emission are respectively $\mathcal{O}(\text{km})$ and $\mathcal{O}(\text{0.3\,km})$÷\cite{price,price-acoustic,rice,attlen}. The applicability of the acoustic detection technique from a physical standpoint is however not so much related to the attenuation length: an inadequately high absolute noise level may definitely impede its applicability in an inappropriate environment. This is still an unknown issue and solutions to remedy have been proposed~\cite{Braun:2010zz}.

\paragraph{Acoustic detection method}
\begin{wrapfigure}{r}{4.5cm}
\centerline{\includegraphics[width=4cm]{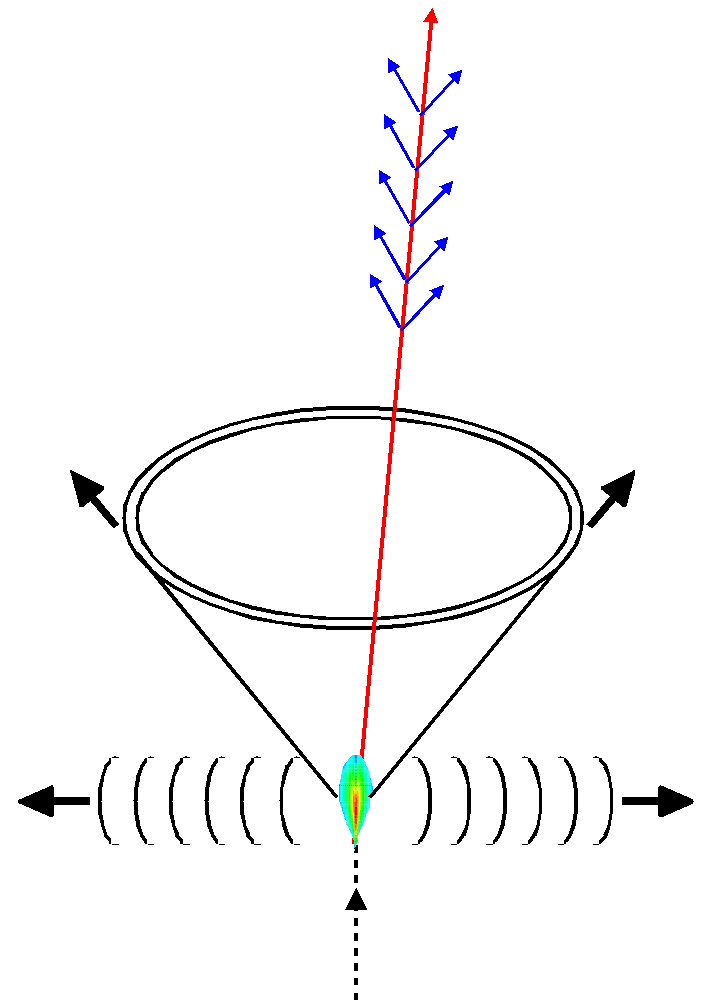}}
\caption{UHE neutrino interaction radio, acoustic and optical signatures.}
\label{fig:RAOsignature} 
\end{wrapfigure}
The emission of a thermo-acoustic pulse following a UHE neutrino interaction is at the origin of the detection technique: the sudden energy release in a medium from the  particles along the shower provokes its thermal expansion with the subsequent shock wave emission. The acoustic pulse is coherent normal to the shower axis and therefore exhibits a "pancake-like" shape as illustrated Fig.~\ref{fig:RAOsignature}. It can be described by means of the wave equation that follows from the equations of motion, continuity and state,
\[
\nabla^2 (p + \frac{1}{\omega_0} \dot{p}) - \frac{1}{c^2}\ddot{p}=-\frac{\beta}{C_p}\frac{\partial E}{\partial t}
\]
which includes a source term on the r.h.s. and where $p$ is the pressure, $\omega_0\equiv \omega_0(f)$ a characteristic attenuation frequency, which depends on medium compressibility and viscosity and frequency (often neglected), $c$ the sound speed, $\beta$ the bulk coefficient of thermal expansion and $C_p$ the specfic heat at constant $p$. The heat deposition is quasi instantaneous, i.e. $\rho_E(\bold{r}',t)=\rho_E(\bold{r'})\delta(t)$ reducing the problem to an integral solution,
\[
p(\bold{r},t)=\int \rho_E(\bold{r}')G(\bold{r}-\bold{r'},t)\drm \bold{r'}
\label{eq:pressure}
\]
with
\[
G(\bold{r},t)=-\frac{\beta}{4\pi C_p} \frac{t-r/c}{r  \sqrt{2\pi}\tau^3} {\rm{e}}^{-\frac{(t-r/c)^2}{2\tau^2}}
\]
where $\tau=\sqrt{r/(\omega_0 c)}$. $G(\bold{r},t)$ is proportional to the pressure pulse generated by a point-like energy deposition $\rho_E(\bold{r})=\rho_0 \delta(\bold{r})$ with an apparent bipolar shape.

In the typical media (salt, ice, water), the peak frequencies of the acoustic pulse are in the range 10-50 kHz, i.e. the signal remains coherent (most of the  energy deposition is within a distance of order of a few cm from the shower axis, corresponding to typical frequencies up to 
$\mathcal{O}(100\,\text{kHz})$.

The efficiency of pressure wave generation is described by the Gr\"uneisen medium parameter $\gamma_g=\beta c^2/ C_p$. The $c$ factor appears when integrating expression~(\ref{eq:pressure}) for a shower-like energy distribution $\rho_E(\bold{r})$ and translates the higher frequencies of the coherent emission, which can be attained. $\gamma_g$ is about 7 times larger in ice than in seawater due to the sound speed difference $c_{\rm{ice}}/c_{\rm{water}}  \approx 2.5$. Ratios of  bulk coefficient of thermal expansion to specific heat are roughly equal for water and ice\footnote{
$c_{\rm{ice}}= 3.92\cdot10^3\,\meterPerSecond,\, c_{\rm{water}}=1.53\cdot10^3 \,\meterPerSecond,\,
\beta_{\rm{ice}}=1.25\cdot 10^{-4}\, \meter^3\, \meter^{-3}\, \kelvin^{-1},\,\beta_{\rm{water}}=2.55\cdot 10^{-4}\, \meter^3\, \meter^{-3}\, \kelvin^{-1}\, C_{p,\rm{ice}}=1.72\cdot 10^3\, {\rm{J}}\,{\rm{kg}}^{-1}\, \kelvin^{-1},\, C_{p,\rm{water}}=3.9\cdot 10^3\, {\rm{J}}\,{\rm{kg}}^{-1}\, \kelvin^{-1}$, see~\cite{price-acoustic}.}. Coincidentally, in ice, 100 kHz is the predicted value above which the ice attenuation length due to scattering at crystal boundaries quickly drops~\cite{price}. The acoustic detection technique in general and its application in seawater is discussed in detail in~\cite{VNiess-thesis}.

Experimentally, the layout of an acoustic array  would would be composed of sparse strings densely (pancake shape) instrumented. In the ice, it would be deployed down to depth smaller than about 1~km, as the absorption length acutely depends on the ice temperature (and the temperature gradient is due to the geothermic flux). 

\paragraph{Radio detection method}
The charge asymmetry in the development of an EM shower and the subsequent coherent radio emission (a short linearly polarized pulse of about a ns) is at the origin of the detection technique.
A shower develops an approx. 20\% charge a symmetry due roughly to the combined effect of shorter pathlengths of the positrons (in-flight anihilation) and of an amplification of the electronic component, via ionization processes (Compton, M{\o}ller, Bhabha),
\begin{enumerate}
\item Shower photons interact with atomic electrons via Compton scattering, the main mechanism for the generation of the charge assymetry,  $\gamma + {\rm{e}}_{\rm{atom}}^{-} \rightarrow \gamma + {\rm{e}}^{-}$
\item Shower positrons interact with atomic electrons via Bhabha scattering,
 ${\rm{e}}^{+}{\rm{e}}_{\rm{atom}}^{-} \rightarrow {\rm{e}}^{+}{\rm{e}}^{-}$.
This process inject new energetic electrons in the shower while reducing the incident positron energy.  
\item Via anihilation in-flight, which terminates an incident positron trajectory,		
 ${\rm{e}}^{+}{\rm{e}}_{\rm{atom}}^{-} \rightarrow \gamma\gamma$
\item Shower electrons interact with atomic electrons via M{\o}ller scattering): 	
 ${\rm{e}}^{-}{\rm{e}}_{\rm{atom}}^{-} \rightarrow {\rm{e}}^{-}{\rm{e}}^{-}$
 \end{enumerate}
All the relativistic particles in the shower contribute to the radio Cherenkov emission (of order of $10^{14}$ particles in a 10 EeV shower). While the Cherenkov emission is faint at radio wavelength, the amplification of the signal is due to the coherence of the collective emission. The coherence compensates the suppression of this emission at long wavelengths and for $L$ of order 10 cm shower core lateral extension. Coherence up to about $\lambda > 4L$ is expected, i.e. up to about GHz frequencies. 

Experimentally, the Cherenkov radio frequency signal is exploited for UHE neutrino detection by operating antennas in the range 0.1 -- 1 GHz. The South Pole ice has adequate properties for the propagation of radio waves at depths down to 1.5 km with an absorption length of 1.2~km (at a few 100's of Mhz).

\section{Conclusions}
From instrumental, phenomenological and methodological standpoints, neutrino astronomy has come a long way since the 60's. In these lectures, we glimpsed at the methodology in a broad sense, trying to establish a link to phenomenology, present and future instrumental techniques.

While the origin of the CRs remains a main focus for neutrino telescopes, the research scope has widened to include questions related to fundamental particle physics and cosmology. Neutrino telescopes offer a rich and promising research program for the next decades.

\acknowledgments
I thank the students and organizers and fellow lecturers for a nice time at the school.

\end{document}